\documentclass[12pt]{article}
\usepackage{amsfonts}
\begin{document}
\newfont{\elevenmib}{cmmib10 scaled\magstep1}%
\renewcommand{\theequation}{\arabic{section}.\arabic{equation}}
\catcode`\@=11
%\makeatletter
\def\smathpalette#1#2{\mathchoice{#1\textstyle{#2}}%
 {#1\scriptstyle{#2}}{#1\scriptscriptstyle{#2}}%
        {#1\scriptscriptstyle{#2}}}
\def\grtrless{\mathrel{\smathpalette\l@ss >}}
\def\l@ss#1#2{\lower-.1pt\vbox{\baselineskip0pt \lineskip-1.2pt %
    \ialign{$\m@th#1\hfil##\hfil$\crcr#2\crcr<\crcr}}}
%\makeatother
\catcode`\@=12

\newcommand{\preprint}{
            \begin{flushleft}
   \elevenmib Yukawa\, Institute\, Kyoto\\
            \end{flushleft}\vspace{-1.3cm}
            \begin{flushright}\normalsize  \sf
            YITP-99-20\\
            KUCP-0132\\
   {\tt hep-th/9907102} \\ July 1999
            \end{flushright}}
\newcommand{\Title}[1]{{\baselineskip=26pt \begin{center}
            \Large   \bf #1 \\ \ \\ \end{center}}}
\newcommand{\Author}{\begin{center}\large \bf
            S.\, P.\, Khastgir$^a$, R.\, Sasaki$^a$
     and K.\, Takasaki$^b$\end{center}}
\newcommand{\Address}{\begin{center} \it
            $^a$ Yukawa Institute for Theoretical Physics, Kyoto
            University,\\ Kyoto 606-8502, Japan \\
     $^b$  Department of Fundamental Sciences,
            Faculty of Integrated Human Studies,\\ Kyoto University,
            Kyoto 606-8501,Japan
      \end{center}}
\newcommand{\Accepted}[1]{\begin{center}{\large \sf #1}\\
            \vspace{1mm}{\small \sf Accepted for Publication}
            \end{center}}
\baselineskip=20pt

\preprint
\thispagestyle{empty}
\bigskip
\bigskip
\bigskip
\Title{Calogero-Moser Models IV: Limits to Toda theory}
\Author

\Address
\vspace{2cm}

\begin{abstract}
Calogero-Moser models  and Toda models
are well-known integrable multi-particle dynamical
systems based on root
systems associated with Lie algebras.
The relation between these two types of integrable models
is investigated at the
levels of the Hamiltonians and the Lax pairs.
The Lax pairs of Calogero-Moser models are specified by t
he representations of
the reflection groups,
which are not the same as those of the corresponding
Lie algebras. The latter specify the Lax pairs of Toda models.
The Hamiltonians of the elliptic
Calogero-Moser models tend to those of Toda models as one of the periods
of the elliptic function goes to infinity, provided the dynamical
variables are properly shifted and the coupling constants are scaled.
On the other hand most of Calogero-Moser Lax pairs, for example,
the root type Lax pairs, do not a have consistent Toda model limit.
The minimal type Lax pairs, which corresponds to the minimal
representations of the Lie algebras, tend to
the Lax pairs of the corresponding Toda models.
\end{abstract}

\newpage
\section{Introduction}
\label{intro}
This is the fifth paper in a series devoted to the integrable dynamical
systems  of
Calogero-Moser type.
In the first three papers \cite{bos} (hereafter referred to as I, II and
III),
various Lax pairs were constructed and the symmetries of the systems were
elucidated.
In the fourth paper \cite{bcs2} a universal Lax pair operator applicable to
the
models based on non-crystallographic as well as crystallographic root
systems was constructed. When suitable representation spaces are chosen, the
universal Lax pair reproduces all the Lax pairs obtained so far and many
other
representations give new Lax pairs. In this paper we focus on the relations
of
Calogero-Moser models and Toda models.

Calogero-Moser models \cite{CalMo} and Toda models \cite{Toda1, Todamod}
are well-known integrable multi-particle dynamical systems based on
the crystallographic root
systems, {\em i.e.} those associated with Lie algebras.
Though they are markedly different at first sight, it has long been known
that
for some particular root systems Toda models can be obtained as a special
limit
of the Calogero-Moser models \cite{Ino1}.
The purpose of this paper is to clarify this limit for all the
Calogero-Moser
models, now that we have a universal framework for the integrable structure
of
these models. The limit problem can be considered at two levels,
or maybe even three.
The first is the limits of the dynamical variables and the Hamiltonian.
The second is those of the Lax pairs.
Since there are many Lax pairs for one and the same Calogero-Moser model,
the
result is expected to be varied. The third level would be the limits of the
solutions of the equations of motion and of the associated linear problem
of the Lax equations. We will not address the problems at the third level in
this paper.

While for the Calogero-Moser models the set of all roots is necessary,
only the simple roots enter the Toda models.
The potentials of Toda models are exponential functions of the dynamical
variables \(q\) (the coordinates), with the mass scale parameters being the
only physically meaningful parameters at the classical level.
The potentials of Calogero-Moser models are more varied. Besides the
(independent) coupling constant(s)
(for the long and short roots in the non-simply laced theories), the
generic elliptic potentials, {\em i.e.}
the Weierstrass \(\wp\) functions have two
primitive periods \(\{2\omega_1,2\omega_3\}\) as adjustable parameters.
The other potentials, the trigonometric, hyperbolic and rational potentials
are obtained as  degenerate cases of the elliptic ones.
As we will show in section two, the Hamiltonian of an elliptic
Calogero-Moser
model based on the root system of Lie algebra \({\mathfrak g}\) tends to the
Hamiltonian of a Toda model based on \({\mathfrak g}\),  its dual
algebra  \({\mathfrak g}^\vee\),  its affine counterpart
\({\mathfrak g}^{(1)}\) or its dual \(({\mathfrak g}^{(1)})^\vee\).
The detailed conditions for the limits will be stated there.
It should be remarked that the significance of the independent coupling
constants in the non-simply laced Calogero-Moser models is somehow
lost in the limit.
The limits of the Lax pairs are more intriguing.

As shown in a previous paper \cite{bcs2}, the essential ingredient of
Calogero-Moser Lax pairs is the representation of the reflection groups,
which are not identical with those of the corresponding Lie algebras.
On the other hand the known Lax pairs of Toda models are formulated in terms
of the representation of the Lie algebras.
This poses an interesting question whether some Toda models Lax pairs which
do not belong to any representations of the corresponding Lie algebras could
be obtained as limits of Calogero-Moser Lax pairs, for example, the root
type
ones.
The answer turns out to be negative as we will show in detail in section
three.
In fact most of Calogero-Moser Lax pairs do not have a consistent Toda model
limit.
The minimal type Lax pairs are shown to have  consistent Toda model limits.

The rest of this paper is organised as follows. In section four we give an
intuitive method for constructing  the Lax pairs of Toda models by explicit
examples. This is inspired by the limit of the minimal type Lax pairs.
Section five is devoted to summary and discussions.
In Appendix \ref{ellipfuns} some definitions and useful formulas of the
elliptic functions are given.
Appendix \ref{othersol} gives some explicit forms of the functions
entering in Calogero-Moser Lax pairs. These functions are also considered in
the limits to Toda models.
Appendix \ref{extwbcr} gives a new Lax pair for the \(BC_r\) Calogero-Moser
model, which is necessary in section three.
The asymptotic forms of various functions appearing in Calogero-Moser Lax
pairs
are given in Appendix \ref{xlimforms}.

\section{From the elliptic potentials to the exponential potentials}
\label{potentials}
\setcounter{equation}{0}

Let us start with the elliptic Calogero-Moser model, which  is a
Hamiltonian system associated with a root system $\Delta$
of rank \(r\).
We consider two types of root systems.
The first type is those root systems associated with finite Lie algebras.
The second is the so-called \(BC_r\) system which is a union of the \(B_r\)
roots and \(C_r\) roots. The dynamical variables are the coordinates
$\{q^{j}\}$ and their canonically conjugate momenta $\{p_{j}\}$, which
will be denoted by vectors in $\mathbf{R}^{r}$
\begin{equation}
  q=(q^{1},\ldots,q^{r}),\qquad p=(p_{1},\ldots,p_{r}).
\end{equation}
The Hamiltonian for the elliptic Calogero-Moser model is
\begin{equation}
   \label{CMHamiltonian}
   \mathcal{H} = {1\over 2} p^{2} + \sum_{\alpha\in\Delta}
   {g_{|\alpha|}^{2}\over |\alpha|^{2}}
   \,V_{|\alpha|}(\alpha\cdot q),
\end{equation}
in which  the potential
functions
$V_{|\alpha|}$ and the real coupling constants $g_{|\alpha|}$ are defined on
orbits of the corresponding
finite reflection group, {\it i.e.} they are
identical for roots in the same orbit.  That is
 \(g_{|\alpha|}=g\) for all roots in simply
laced models and  \(g_{|\alpha|}=g_L\)
for the long roots and \(g_{|\alpha|}=g_S\) for
the short roots in
non-simply laced models.
The potential
functions
$V_{|\alpha|}$ are:
\begin{enumerate}
\item {\em Untwisted elliptic potential\/}. This applies to all of the
 root systems associated with Lie algebras and the potential function is
\begin{equation}
   V_L(\alpha\cdot q)=V_S(\alpha\cdot q)=
   \wp(\alpha\cdot q|\{2\omega_1,2\omega_3\}),
   \label{simppot}
\end{equation}
in which \(\wp\) is the Weierstrass \(\wp\) function with a pair
of primitive periods
\(\{2\omega_1,2\omega_3\}\) (\ref{periods}).
Throughout this paper we adopt the convention that
the Weierstrass \(\wp\), \(\zeta\), and \(\sigma\) functions have the above
standard periods, unless otherwise stated.
\item {\em Twisted elliptic potential\/}. This applies to all of the
non-simply laced root systems. Except for the \(G_2\) model,
the potential functions are
\begin{equation}
   V_L(\alpha\cdot q)=
   \wp(\alpha\cdot q|\{2\omega_1,2\omega_3\}),
   \qquad
   V_S(\alpha\cdot q)= \wp(\alpha\cdot q|\{\omega_1,2\omega_3\}).
\end{equation}
That is, the potential for the short roots has one half of the standard
period
in one direction, which we choose to be \(\omega_1\). For the \(G_2\)
model,
\begin{equation}
   V_L(\alpha\cdot q)=
  \wp(\alpha\cdot q|\{2\omega_1,2\omega_3\}),
   \qquad  V_S(\alpha\cdot q)=
   \wp(\alpha\cdot q|\{{2\omega_1\over3},2\omega_3\}).
\end{equation}
Derivation of the twisted models from the untwisted ones by folding
is given in paper II, \cite{bos}.
\item {\em Untwisted and twisted potentials for the \(BC_r\) system. }
The
\(BC_r\) root system consists of the long,  middle and  short roots,
\(\Delta=\Delta_L\cup\Delta_M\cup\Delta_S\).
The untwisted model has the same potential for all the roots
\begin{equation}
   V_L(\alpha\cdot q)=V_M(\alpha\cdot q)=V_S(\alpha\cdot q)=
   \wp(\alpha\cdot q|\{2\omega_1,2\omega_3\}).
   \label{simppot2}
\end{equation}
The twisted model has potentials with the full, a half and
a fourth periods:
\begin{eqnarray}
 V_L(\alpha\cdot q)&=&
   \wp(\alpha\cdot q|\{2\omega_1,2\omega_3\}),
   \qquad
   V_M(\alpha\cdot q)= \wp(\alpha\cdot q|\{\omega_1,2\omega_3\}),
   \nonumber\\
  V_S(\alpha\cdot q)&=& \wp(\alpha\cdot q|\{\omega_1/2,2\omega_3\}).
\end{eqnarray}
There are three independent coupling constants \(g_L\),
\(g_M\) and \(g_S\) in both cases.

In the discussion below the root systems associated with the Lie algebras
are assumed. Modification for the \(BC_r\) case is straightforward.
\end{enumerate}

On taking the limits to Toda models it is convenient to  adopt the following
parametrisation of the periods:
\begin{equation}
   \omega_1=-i\pi, \quad
   \omega_3 \in{\bf R}_{+},\quad
   \tau\equiv{\omega_{3}\over{\omega_{1}}}=i\omega_{3}/\pi.
   \label{hermcond}
\end{equation}
Then the above Hamiltonian (\ref{CMHamiltonian}) is
real  for real dynamical variables \(p,q\)
and coupling constants
\(g_{|\alpha|}\).
If we let \(\omega_3\to+\infty\) for fixed \(u\) the elliptic
potentials tend to
the hyperbolic ones
\begin{equation}
V_L(u)={1\over{12}}+{1\over4}{1\over{\sinh^2{u/2}}},\quad
V_S(u)={1\over{3}}+{1\over{\sinh^2{u}}}.
\label{potlim}
\end{equation}

In order to obtain the exponential potential from the elliptic
potentials  we  follow the general prescription as explained in the Appendix
(\ref{shiftu})--(\ref{g1wpscal}), \cite{Ino1,DHPh2}. First we
shift the dynamical variable \(q\)
\begin{equation}
   q=Q-2\omega_3\delta\,v,\quad v\in{\bf R}^r,
   \label{qshift}
\end{equation}
in which \(\delta\) is a positive parameter and \(v\) is an as yet
unspecified vector in \({\bf R}^r\). Let us require that \(v\) has a
non-vanishing scalar product with all the roots in \(\Delta\)
\begin{equation}
   \alpha\cdot v\neq 0,\quad \forall\alpha\in\Delta.
   \label{nonvan}
\end{equation}
Suppose there are some roots which are orthogonal to \(v\).
They form a sub-root system of \(\Delta\).
The potential functions for such roots will
tend to the hyperbolic potentials
(\ref{potlim}) as \(\omega_3\to+\infty\), since their arguments
\(\alpha\cdot
q=\alpha\cdot Q\) are fixed. This justifies the above requirement
(\ref{nonvan}).
It should be remarked that the above shift (\ref{qshift})
breaks the Weyl invariance of the Calogero-Moser models, by the introduction
of the special vector \(v\).
On the other hand the set of positive roots \(\Delta_+\) and consequently
the set of simple roots \(\Pi\) can
be defined in terms of \(v\):
\begin{equation}
   \Delta_+=\{\alpha\in\Delta,\ \alpha\cdot v>0\},
   \qquad \Pi:\mbox{set of simple
   roots}.
   \label{posrt}
\end{equation}
Because of the \(2\omega_3\) periodicity and the even parity of the
potentials,
\(V_{|\alpha|}(u)=V_{|\alpha|}(-u)\) we can assume,
without loss of generality, that
\begin{equation}
   \max_{\alpha\in\Delta_+}\delta\alpha\cdot v<1.
\end{equation}
In fact we require that \(\delta\) should satisfy a stronger condition
\begin{equation}
   \delta\,((\alpha\cdot v)_{\mbox{min}}+(\alpha\cdot v)_{\mbox{max}})\leq1,
   \label{deltaeq}
\end{equation}
which implies that
\begin{equation}
   \delta\,(\alpha\cdot v)_{\mbox{max}}\leq 1-\delta\,(\alpha\cdot
   v)_{\mbox{min}}
\end{equation}
and
\begin{equation}
   2\delta\,(\alpha\cdot v)_{\mbox{min}}\leq1.
\end{equation}

By comparing the shift formula
\[
   \alpha\cdot q=\alpha\cdot Q-2\omega_3\delta\,\alpha\cdot v,
\]
with the limit formula of the potentials (\ref{shiftu})--(\ref{g1wpscal}),
we find that for the following scalings of the coupling constants
\begin{equation}
   g_L=m_L\,e^{\omega_3\delta|\alpha\cdot v|_{\mbox{\scriptsize min}}},\quad
   g_S=\left\{
   \begin{array}{ll}
      m_S\,e^{\omega_3\delta|\alpha\cdot
      v|_{\mbox{\scriptsize min}}}/\sqrt2,&
      \mbox{untwisted potential},\\[6pt]
      m_S\,e^{2\omega_3\delta|\alpha\cdot
      v|_{\mbox{\scriptsize min}}}/2\sqrt2,&
      \mbox{twisted potential},
   \end{array}
   \right.
\end{equation}
the elliptic potentials {\em vanish} for all \(\alpha\cdot q\) except
for those roots having the minimum (and the maximum) value of the scalar
product with the fixed vector \(v\), for which the exponential potentials
are
obtained. That is we have
\begin{equation}
   g^2_LV_L(\alpha\cdot q)
   {{}\atop{\longrightarrow\atop{ \omega_3\to+\infty}}}
   \left\{
   \begin{array}{cl}
      m^2_L\,e^{\alpha\cdot Q} &\mbox{for such \(\alpha\in\Delta_+\) that
      \(\alpha\cdot v\) is minimum,}\\[6pt]
      m^2_L\,e^{-\alpha\cdot Q} &\mbox{for such \(\alpha\in\Delta_+\) that
      \(\alpha\cdot v\) is maximum,}\\[6pt]
      0&\mbox{otherwise,}
   \end{array}
   \right.
   \label{genlim}
\end{equation}
and the corresponding formula for \(g^2_SV_S(\alpha\cdot q)\).
It should be noted that for the twisted models the minimum (maximum) of
\((\alpha\cdot v)\) can be different for \(V_L\) and \(V_S\), since only
the long (short) roots contribute to \(V_L\) (\(V_S\)).
However, for the second possibility (\(\alpha\cdot v\) is maximum) to occur,
the parameter \(\delta\) must be so chosen as to saturate the inequality in
(\ref{deltaeq}).
As we will see shortly, (\ref{satl}), (\ref{unsats}),
the saturation occurs only for the
long root potentials.

In the  formula (\ref{genlim}) we considered only the positive roots
\(\alpha\).
The number of non-vanishing potential terms of the resulting theory
is determined by those positive roots which give the minimum (and the
maximum)
of \(\alpha\cdot v\).
Since all the positive roots are linear combination of simple roots with
non-negative integer coefficients, the minimum can be attained by the simple
roots only and the maximum by the highest root.
There are a maximal number of potential terms when the minimum is attained
by
all the simple roots. All the other cases can be considered as arising from
Calogero-Moser models based on some sub-root system of \(\Delta\).

\subsection{Models based on the root systems of Lie algebras}

First let us discuss the models based on root systems associated with finite
Lie algebras. We consider the case that the minimum of  \(\alpha\cdot
v\) is attained by all the simple roots.
We adopt the convention that the long roots have squared length 2,
\(\alpha_L^2=2\).
It is well-known that the Weyl vector \(\rho\) and its dual \(\rho^\vee\)
defined by
\begin{equation}
   \rho={1\over2}\sum_{\alpha\in\Delta_+}\alpha,\qquad
   \rho^\vee={1\over2}\sum_{\alpha\in\Delta_+}\alpha^\vee,\qquad
   \alpha^\vee=2\alpha/\alpha^2,
\end{equation}
satisfy the above criterion. In fact we have
\begin{equation}
   \rho\cdot\alpha_i={\alpha_i^2\over2},\qquad
   \rho^\vee\cdot\alpha_i=1,\qquad \forall\alpha_i\in\Pi,
\end{equation}
and
\begin{equation}
   \rho\cdot\alpha_h=h^\vee-1,\quad
   \rho^\vee\cdot\alpha_h=h-1,\quad \alpha_h:\ \mbox{highest root},
\end{equation}
in which \(h\) and \(h^\vee\) are the Coxeter number and the dual Coxeter
number, respectively. For the choice \(v=\rho^\vee\), the minimum is always
1
for the long and short simple roots.
For the non-simply laced root system, there are two different values
of minimum for \(v=\rho\), the Weyl vector,
1 for the long roots and \(1/2\) for
the short roots. This corresponds to the existence of two different coupling
constants \(g_L\) and \(g_S\).
For all  root systems \(\Delta\) except
\(A_1\), we have
\(|\rho\cdot\alpha|_{\mbox{min}}<|\rho\cdot\alpha|_{\mbox{max}}\), with the
long roots satisfying
\begin{equation}
   |\rho\cdot\alpha|_{\mbox{min}}+
   |\rho\cdot\alpha|_{\mbox{max}}=h^\vee,\qquad
   |\rho^\vee\cdot\alpha|_{\mbox{min}}+
   |\rho^\vee\cdot\alpha|_{\mbox{max}}=h
   \label{satl}
\end{equation}
and  the short roots
\begin{equation}
   |\rho\cdot\alpha|_{\mbox{min}}+
   |\rho\cdot\alpha|_{\mbox{max}}<h^\vee,\qquad
   |\rho^\vee\cdot\alpha|_{\mbox{min}}
   +|\rho^\vee\cdot\alpha|_{\mbox{max}}<h.
   \label{unsats}
\end{equation}
That is, the saturation of of the inequality (\ref{deltaeq})
occurs only for the long roots for the choices of
\(\delta=1/h\) for \(v=\rho^\vee\) or \(\delta=1/h^\vee\) for \(v=\rho\).
For \(A_1\) we have
\begin{equation}
   \rho=\rho^\vee,\qquad
   |\rho\cdot\alpha|_{\mbox{min}}=
   |\rho\cdot\alpha|_{\mbox{max}}=1, \quad h=2.
   \label{a1}
\end{equation}

\bigskip
In the Hamiltonian of an elliptic
Calogero-Moser model based on a root system \(\Delta\) which is associated
with a Lie algebra ${\mathfrak g}$,
we redefine the dynamical variables from \(\{p,q\}\) to \(\{P,Q\}\)
\begin{equation}
   q=Q-2\omega_3\delta\,v,\quad p=P,\quad v=\rho\quad\mbox{or}\quad
   \rho^\vee,
   \label{qshift2}
\end{equation}
and take the limit \(\omega_3\to+\infty\) with \(\omega_1=-i\pi\).
The coupling constants are also scaled:
\begin{equation}
   g_L=m_L\,e^{\omega_3\delta},\quad
   \begin{array}{llll}
   g_S&=\left\{
   \begin{array}{lc}
      m_S\,e^{\omega_3\delta}/\sqrt2,& \mbox{untwisted
      potential},\\[6pt]
      m_S\,e^{2\omega_3\delta}/2\sqrt2,& \mbox{twisted potential},
      \end{array}
      \right\}& \mbox{for}& v=\rho^\vee,
      \\[18pt]
      g_S&=\left\{
      \begin{array}{lc}
      m_S\,e^{\omega_3\delta/2}/\sqrt2,& \mbox{untwisted
      potential},\\[6pt]
      m_S\,e^{\omega_3\delta}/2\sqrt2,& \mbox{twisted potential},
      \end{array}
      \right\}& \mbox{for}& v=\rho.
   \end{array}
   \label{coupscal}
\end{equation}
In this limit we arrive at the Hamiltonian of the Toda models associated
with the Lie algebra ${\mathfrak g}$,
its dual algebra ${\mathfrak g}^\vee$,
the untwisted affine Lie algebra
${\mathfrak g}^{(1)}$ or its dual $({\mathfrak g}^{(1)})^\vee$
depending on the types of the potential,
untwisted or twisted, and the values
of the parameter \(\delta\).

For the simply laced root system \(\Delta\), we have the Toda system
associated with the Lie algebra ${\mathfrak g}$ for \(\delta<1/h\):
\begin{equation}
   {\cal H}={1\over2}P^2+m^2\sum_{\alpha_i\in\Pi}e^{\alpha_i\cdot Q},
\end{equation}
and the  Toda system
associated with the untwisted affine Lie algebra ${\mathfrak g}^{(1)}$ for
\(\delta=1/h\):
\begin{equation}
   {\cal H}={1\over2}P^2+m^2\left(\sum_{\alpha_i\in\Pi}e^{\alpha_i\cdot Q}
   +e^{\alpha_0\cdot Q}\right),
   \label{untwaftoda}
\end{equation}
in which \(\alpha_0\) is the affine root  \(\alpha_0=-\alpha_h\).
For the \(A_1\) case with \(\delta=1/2\)
the extreme situations (\ref{a1}) and
(\ref{g1wpscal}) apply and give the above result (\ref{untwaftoda}).

For a non-simply laced root system \(\Delta\) and an untwisted potential,
we have the Toda system associated with the Lie algebra ${\mathfrak g}$:
\begin{equation}
   {\cal H}={1\over2}P^2+m^2_L
   \sum_{\alpha_i\in\Pi\cap\Delta_L}e^{\alpha_i\cdot
   Q}+
   m^2_S\sum_{\alpha_i\in\Pi\cap\Delta_S}e^{\alpha_i\cdot
   Q},
\end{equation}
for \(v=\rho^\vee\) and \(\delta<1/h\) or for \(v=\rho\) and
\(\delta<1/h^\vee\), and the  Toda system
associated with the untwisted affine Lie algebra ${\mathfrak g}^{(1)}$:
\begin{equation}
   {\cal
   H}={1\over2}P^2+m^2_L\left(\sum_{\alpha_i\in\Pi\cap\Delta_L}
   e^{\alpha_i\cdot
   Q} +e^{\alpha_0\cdot Q}\right)+
   m^2_S\sum_{\alpha_i\in\Pi\cap\Delta_S}e^{\alpha_i\cdot Q},
\end{equation}
for \(v=\rho^\vee\)  and \(\delta=1/h\) or for \(v=\rho\)  and
\(\delta=1/h^\vee\).
In these formulas \(\Delta_L\) (\(\Delta_S\))
is the set of long (short) roots.

For a non-simply laced root system \(\Delta\) and a twisted potential,
we have the Toda system associated with the dual Lie algebra ${\mathfrak
g}^\vee$:
\begin{equation}
   {\cal H}={1\over2}P^2+m^2_L\sum_{\alpha_i\in\Pi\cap\Delta_L}
   e^{\alpha_i\cdot
   Q}+
   m^2_S\sum_{\alpha_i\in\Pi\cap\Delta_S}e^{2\alpha_i\cdot
   Q},
   \label{twtoda}
\end{equation}
for \(\delta<1/h\) and \(v=\rho^\vee\) or  \(\delta<1/h^\vee\) and
\(v=\rho\),
and the  Toda system
associated with  the twisted affine Lie algebra
$({\mathfrak g}^{(1)})^\vee$:
\begin{equation}
   {\cal
   H}={1\over2}P^2+m^2_L\left(\sum_{\alpha_i\in\Pi\cap\Delta_L}
   e^{\alpha_i\cdot
   Q} +e^{\alpha_0\cdot Q}\right)+
   m^2_S\sum_{\alpha_i\in\Pi\cap\Delta_S}e^{2\alpha_i\cdot Q},
   \label{twaftoda}
\end{equation}
for \(v=\rho^\vee\) and \(\delta=1/h\) or for \(v=\rho\)  and
\(\delta=1/h^\vee\).
The formulas for the twisted potential cases (\ref{twtoda}) and
(\ref{twaftoda}) are valid for all the non-simply laced models
except for the one based on \(G_2\). In this case the last term in
(\ref{twtoda}) and (\ref{twaftoda}) should be changed to
\(e^{3\alpha_i\cdot
Q}\) with appropriate scaling of \(g_S\).

A few remarks are in order. In the Calogero-Moser models the meaning of the
coupling constant is clear.
They specify the strength of the repulsive potentials near the boundary of
the
Weyl chambers.
Thus the independence of the coupling constants is quite crucial in the
Calogero-Moser model.
These properties are lost in the transition to the Toda models by the shift
of
the dynamical variables (\ref{qshift2}) and the scalings of the coupling
constants (\ref{coupscal}).
The remaining parameters \(m_L\) and \(m_S\), (\ref{coupscal}), are
generally
considered as giving mass scales of the Toda theories.
For the Toda theories based on finite Lie algebras, this interpretation
is not adequate, since these theories are conformally invariant. In fact,
in the Toda theories based on finite Lie algebras \({\mathfrak g}\), the
mass
parameters \(m_i^2\) can be changed arbitrarily.
Suppose we start from the Hamiltonian
\[
{\cal H}={1\over2}P^2+\sum_{\alpha_i\in\Pi}m_i^2e^{\alpha_i\cdot
Q}
\]
and make a shift
\begin{equation}
   Q=Q'+\sum_{\alpha_i\in\Pi}{2\over{\alpha_i^2}}\lambda_i
   \log({{m_i^\prime}^2\over{m_i^2}}),
   \label{coupredef}
\end{equation}
in which \(\{\lambda_i\}\), \(i=1,\ldots,r\) are the fundamental weights,
satisfying \(\lambda_i\cdot\alpha_j^\vee=\delta_{ij}\)
for \(\alpha_j\in\Pi\).
We arrive at
\begin{equation}
   {\cal H}={1\over2}P^2+\sum_{\alpha_i\in\Pi}
   {m_i^\prime}^2e^{\alpha_i\cdot
   Q^\prime}.
   \label{redefdham}
\end{equation}

\subsection{\(BC_r\) model}
The \(BC_r\) root system consists of three parts,
long,  middle and short roots:
\[
   \Delta_{BC_r}=\Delta_L\cup\Delta_M\cup\Delta_S,
\]
in which the roots are conveniently expressed in terms of
an orthonormal basis of \({\bf R}^r\):
\begin{equation}
   \Delta_L=\{\pm 2e_j\},\quad
   \Delta_M=
   \{\pm e_j\pm e_k\},  \quad
   \Delta_S=
   \{\pm e_j\}: \quad
   j=1,\ldots,r.
   \label{bcnroots}
\end{equation}
The set of simple roots is the same as that of \(B_r\):
\begin{equation}
   \Pi=\{e_r\}\cup\{e_j-e_{j+1},\ j=1,\ldots,r-1\}.
   \label{bcrsimp}
\end{equation}
If we define
\begin{equation}
   \rho^\vee=\sum_{j=1}^r(r+1-j)e_j,\quad h=2r+1,
   \label{bcrrho}
\end{equation}
and the following scalings
\begin{equation}
   g_L=\sqrt2m_L\,e^{\omega_3\delta},\quad
   g_M=
   {m_M}\,e^{\omega_3\delta},\quad
   g_S=
   {m_S}\,e^{\omega_3\delta}/\sqrt2,
   \label{gbcruntw}
\end{equation}
for the untwisted potential we obtain the non-affine \(B_r\) Toda model for
\(\delta<1/h\) and \(A_{2n}^{(2)}\) Toda model for \(\delta=1/h\):
\begin{equation}
   {\cal H}={1\over2}P^2+m_L^2\,e^{-2Q_1}
   +m_M^2\sum_{j=1}^{r-1}e^{Q_j-Q_{j+1}}+
   m_S^2\,e^{Q_r}.
\end{equation}
For the following scalings
\begin{equation}
   g_L=\sqrt2m_L\,e^{\omega_3\delta},\quad
   g_M=
   {m_M}\,e^{2\omega_3\delta}/2,\quad
   g_S=
   {m_S}\,e^{4\omega_3\delta}/4\sqrt2,
   \label{gbcrtw}
\end{equation}
and the twisted potentials,
we obtain  the non-affine \(C_r\) Toda model for
\(\delta<1/h\) and another form of the \(A_{2n}^{(2)}\) Toda model for
\(\delta=1/h\):
\begin{equation}
   {\cal H}={1\over2}P^2+m_L^2\,e^{-2Q_1}
   +m_M^2\sum_{j=1}^{r-1}e^{2(Q_j-Q_{j+1})}+
   m_S^2\,e^{4Q_r}.
\end{equation}
this is due to the fact that the \(A_{2n}^{(2)}\) root system is self-dual.
%%%%%%%%%%%%%%%%%%%%%%%%%%%%%%%
\section{Limits of the Lax pairs}
\setcounter{equation}{0}
In this section we discuss the corresponding limits of the Calogero-Moser
Lax pairs. Contrary to the Hamiltonian case which has well-defined limits to
the Toda model Hamiltonians, the limits of the Lax pairs are diverse, some
having well-defined limits to the Toda model Lax pairs and some not.
We consider two different types of Calogero-Moser Lax pairs,
the root type and the minimal type \cite{bos}, both expressing the canonical
equations of motion in terms of a pair of matrices \(L\) and \(M\)
\begin{equation}
    \dot{L}={d\over{dt}}L=[L,M].
    \label{eq:laxeq0}
\end{equation}
Roughly speaking the elements of the \(L\) matrix are square roots
of the Hamiltonian, since Tr\((L^2)\propto {\cal H}\).
These Lax pairs depend on an additional parameter \(\xi\), the spectral
parameter, which requires a shift proportional to \(\omega_3\) as we will
see
presently.
Therefore the existence of a limit of the Hamiltonian does not imply that of
a Lax pair, not to mention those of both types, since they correspond to
different types of square roots of the Hamiltonian.
In the following we examine the limits of various Lax pairs in turn.

\subsection{Minimal type Lax pair for simply laced root systems}

This type of Lax pairs has the simplest structure and has a well-defined
limit
to the Lax pair of the corresponding Toda model, as expected.
This applies to the models based on the root systems of \(A_r\), \(D_r\),
\(E_6\) and \(E_7\).

The minimal type Lax pairs have the following form,
\begin{eqnarray}
   L(q,p,\xi) &=&p\cdot H +X, \nonumber\\
   M(q,\xi)&=&D +Y.
   \label{laxpair}
\end{eqnarray}
The matrix elements of \(L\) and \(M\) are labelled by the weights
of a minimal representation,  (I.4.1) in \cite{bos}.
The matrices  \(H\) and \(D\) are diagonal
\begin{equation}
   H_{\mu\nu}=\mu\delta_{\mu\nu} ~~~{\rm and}~~~
   D_{\mu\nu}=\delta_{\mu\nu}
   D_{\mu},~~~D_{\mu}=ig\sum_{\Delta \ni
   \beta=\mu-\nu}\wp(\beta\cdot q).
   \label{diag}
\end{equation}
The matrices \(X\) and \(Y\) have the form
\begin{equation}
    X=ig\sum_{\alpha\in\Delta}x(\alpha\cdot
    q, \xi)E(\alpha),\quad
    Y=ig\sum_{\alpha\in\Delta}y(\alpha\cdot q, \xi)E(\alpha),
    \label{minXYdef}
\end{equation}
in which \(\xi\) is the spectral parameter and
$E(\alpha)_{\mu\nu}=\delta_{\mu-\nu,\alpha}$.
It should be stressed
that the $H$ and $E(\alpha)$ here are {\em not} the Lie algebra generators
for
the associated Lie algebra \({\mathfrak g}\), though they satisfy relations
\begin{eqnarray}
    [H,E(\alpha)]&=&\alpha E(\alpha),\quad
   \ [H,[E(\alpha),E(\beta)]]=(\alpha+\beta)[E(\alpha),E(\beta)],
   \nonumber\\
   \ E(-\alpha)&=&E(\alpha)^T,\quad
   [E(\alpha),E(-\alpha)]=\alpha\cdot H.
    \label{eq:algebra}
\end{eqnarray}
The function \(x(u,\xi)\) ( \(y(u,\xi)={\partial_u}x(u,\xi)\)) is
a solution of a certain functional equation involving the potential function
\cite{OP1}, (I.2.14), \cite{bos} and
it factorises the potential as
\begin{equation}
   x(u,\xi)x(-u,\xi)=-\wp(u)+\wp(\xi).
   \label{fac}
\end{equation}
It is not unique in the sense that if \(x(u,\xi)\) is a solution then
\begin{equation}
   \tilde{x}(u,\xi)=x(u,\xi)\,e^{b(\xi)u},
   \quad b(\xi):\ \mbox{an arbitrary
   function}
   \label{gauge}
\end{equation}
is also a solution providing another factorisation, (II.2.27), \cite{bos}.
As shown in the previous section, for
\begin{equation}
   q=Q-2\omega_3\delta\,\rho,\quad g=m\,e^{\omega_3\delta},
   \label{qg-scale}
\end{equation}
we have the following limits of the potential
\begin{equation}
   g^2\wp(\alpha\cdot q)
   {{}\atop{\longrightarrow\atop{\omega_3\to+\infty}}}
   \left\{
   \begin{array}{cll}
      m^2\,e^{\alpha\cdot Q} &\alpha\in\Pi,
      &\delta<1/h,\\[6pt]
      m^2\,e^{-\alpha_h\cdot Q} &\alpha_h: \mbox{highest root},
      &\delta=1/h,\\[6pt]
      0&\mbox{otherwise.}&
   \end{array}
   \right.
   \label{genlim2}
\end{equation}
(In a simply laced root system \(\rho=\rho^\vee\).)
In order to have a finite limit for \(g^2\wp(\xi)\), \(\xi\) needs to be
shifted, too:
\begin{equation}
   \xi=\log Z-2\omega_3\epsilon,\quad Z\in{\bf R}_+,\quad
    \delta<\epsilon\leq1/2,
   \label{xiscale}
\end{equation}
in which the \(\omega_3\)-independent part of \(\xi\) is parametrised by a
positive number \(Z\) for later convenience. From (\ref{wpappr}) we then
find
that
\(g^2\wp(\xi)\propto e^{2\omega_3(\delta-|\epsilon|)}\) has a finite limit
for
\(\omega_3\to+\infty\)  up to a diverging
constant. This constant is canceled by the one coming
from \(g^2\wp(\alpha\cdot
q)\) in the factorisation formula (\ref{fac}).
For the consistency of the limit of the Lax pair with that of the
Hamiltonian,
the following limit of \(gx(\alpha\cdot q,\xi)\)
\begin{equation}
   gx(\alpha\cdot q,\xi)\to
   \left\{
   \begin{array}{cl}
      \mbox{finite},&
      \mbox{for}
      \quad
      \pm\alpha_i\in\Pi\quad (\delta\leq1/h)\quad \mbox{and}\
      \pm\alpha_h\quad (\delta=1/h),\\[6pt]
      0,&\mbox{otherwise,}
   \end{array}
   \right.
   \label{xcond}
\end{equation}
is necessary. It is obvious that this condition selects, if any, a unique
solution among the equivalent ones related by the symmetry transformation
(\ref{gauge}).

We will show that the following solution \cite{Krichever}
\begin{equation}
   x(u,\xi)={\sigma(\xi-u)\over{\sigma(\xi)\sigma(u)}}
   \exp({\zeta(\xi) u})
   \label{funx}
\end{equation}
satisfies the above condition (\ref{xcond}) and gives a minimal type Lax
pair
which tends to a Toda model Lax pair in the limit \(\omega_3\to+\infty\).
The asymptotic form of this solution can be evaluated  using formulas
(\ref{sigapp}) and (\ref{sigapp2})
\footnote{
The formulas (\ref{limitgx}), (\ref{limitgx1}), (\ref{limitgx2}) and
(\ref{limitgx3}) are valid for all the root systems except for \(A_1\). The
\(A_1\) case needs be considered separately because of (\ref{a1}).
The Toda Lax pair (\ref{todalax}), (\ref{todalax1})is valid for all the
cases
including the
\(A_1\).}. For
\(\sigma(\xi)\) only (\ref{sigapp}) is needed and for \(\sigma(\alpha\cdot
q)\) and
\(\sigma(\xi-\alpha\cdot q)\) both (\ref{sigapp}) and (\ref{sigapp2}) are
necessary according to the range of the arguments.

\noindent
For positive roots \(\alpha\):
\begin{eqnarray}
   gx(\alpha\cdot q,\xi)\rightarrow&-m\exp({{\alpha\cdot Q}\over{2}})
   \exp[\omega_3\delta(1-
   \rho\cdot\alpha)], ~~~~~~~~~~~~~~~~
   0<\epsilon -\delta\rho\cdot\alpha<1,
   \nonumber\\
   \rightarrow&{mZ}\exp(-{{\alpha\cdot
   Q}\over{2}})\exp[\omega_3(\delta+\delta
   \rho\cdot\alpha-2\epsilon)],~~~~
   -1<\epsilon -\delta\rho\cdot\alpha\leq0,
   \label{limitgx}
\end{eqnarray}
whilst for negative roots \(\alpha\):
\begin{eqnarray}
   gx(\alpha\cdot q,\xi)\rightarrow&m\exp(-{{\alpha\cdot
   Q}\over{2}})\exp[\omega_3\delta(1+
   \rho\cdot\alpha)], \qquad\qquad
   0<\epsilon -\delta \rho\cdot\alpha<1,\nonumber\\
   \rightarrow&-{m\over{Z}}\exp({{\alpha\cdot
   Q}\over{2}})\exp[\omega_3(2\epsilon+\delta
   -\delta\rho\cdot\alpha-2)]~~~~~
   1\leq\epsilon -\delta \rho\cdot\alpha<2.
   \label{limitgx1}
\end{eqnarray}

For example, for the parameter ranges
\begin{equation}
   \delta(h-1)<\epsilon\leq1/2,\qquad \mbox{or}\quad \delta<1/h,
   \quad \epsilon=1/2,
   \label{simprange}
\end{equation}
the function \(gx(\alpha\cdot q,\xi)\) is non-vanishing only for
the positive and
negative simple roots. This corresponds to the Toda models based on simply
laced finite Lie algebras.
For
\begin{equation}
   \epsilon=1/2,\quad \delta=1/h
\end{equation}
the function \(gx(\alpha\cdot q,\xi)\) is non-vanishing only for the
positive
and negative simple roots and the highest roots.
This corresponds to the Toda models based on simply laced affine Lie
algebras.

In the other Lax matrix  $M$, (\ref{laxpair}), there are terms $D$ and $Y$.
The limit of
$D$ simply vanishes under the scaling because it contains the term
$g\wp(\alpha\cdot q)$, which has a power of $g$ less than that
appearing in the potential, which has a finite limit. Calculating the limit
for $Y$ is also not very difficult as we have already  calculated the
limit of $gx(\alpha\cdot q)$ and $gy(\alpha\cdot q)=gx'(\alpha\cdot q)$.
The result is summarised as follows
( \(\alpha\) is a positive root):
\begin{equation}
   \begin{array}{llll}
      gx(\alpha\cdot q,\xi)&\rightarrow&-m\exp({{\alpha\cdot Q}\over{2}})&
      \mbox{simple roots}\\[6pt]
      &\rightarrow&{mZ}\exp(-{{\alpha\cdot
      Q}\over{2}})&\mbox{highest root}\\[6pt]
      &\rightarrow&0&\mbox{otherwise}\\[8pt]
      gy(\alpha\cdot q,\xi)&\rightarrow&-{m\over2}
      \exp({{\alpha\cdot Q}\over{2}})&
      \mbox{simple roots}\\[6pt]
      &\rightarrow&-{mZ\over2}\exp(-{{\alpha\cdot
      Q}\over{2}})&\mbox{highest root}\\[6pt]
      &\rightarrow&0&\mbox{otherwise}
   \end{array}
   \label{limitgx2}
\end{equation}
and
\begin{equation}
   \begin{array}{llll}
      gx(-\alpha\cdot q,\xi)&\rightarrow&
      m\exp({{\alpha\cdot Q}\over{2}})&
      \mbox{simple roots}\\[6pt]
      &\rightarrow&-{m\over Z}\exp(-{{\alpha\cdot
      Q}\over{2}})&\mbox{highest root}\\[6pt]
      &\rightarrow&0&\mbox{otherwise}\\[8pt]
      gy(-\alpha\cdot q,\xi)&\rightarrow&
      -{m\over2}\exp({{\alpha\cdot Q}\over{2}})&
      \mbox{simple roots}\\[6pt]
      &\rightarrow&-{m\over{2Z}}\exp(-{{\alpha\cdot
      Q}\over{2}})&\mbox{highest root}\\[6pt]
      &\rightarrow&0&\mbox{otherwise.}
   \end{array}
   \label{limitgx3}
\end{equation}
The Lax pair now reads
\begin{equation}
   L=P\cdot H-im\sum_{\alpha \in \Pi}\exp({{\alpha\cdot
   Q}\over 2})[E(\alpha)-E(-\alpha)] +im\exp({{\alpha_0\cdot
   Q}\over 2})[ZE(-\alpha_0)-Z^{-1}E(\alpha_0)],
   \label{todalax}
\end{equation}
\begin{equation}
   M=-{i\over 2}m\sum_{\alpha
   \in\Pi}\exp({{\alpha\cdot Q}\over 2})[E(\alpha)+E(-\alpha)] -{i\over
   2}m\exp({{\alpha_0\cdot Q}\over 2})[ZE(-\alpha_0)+Z^{-1}E(\alpha_0)].
   \label{todalax1}
\end{equation}
For the Toda models based on a finite Lie algebra
\({\mathfrak g}\), one should
drop the terms containing the affine root \(\alpha_0\).
The parameter \(Z\) which is a scaled version of
the original spectral parameter
\(\xi\) now plays the role of a spectral parameter for the affine Toda model
based on \({\mathfrak g}^{(1)}\).
It should be stressed again that although the  matrices \(E(\alpha)\) are
not
Lie algebra generators as a whole,  they satisfy the necessary relations
for the Toda model Lax pairs
\begin{equation}
   [E(\alpha),E(-\beta)]=0,\quad \alpha,\beta\in\Pi\cup \{\alpha_0\},
\end{equation}
on top of those listed in (\ref{eq:algebra}).
The non-vanishing intermediate state \(\kappa\) in the above commutation
relations either does not exist (for the case
\(\alpha\cdot \beta=-1\)), or if it exists it forms a pair which cancels
with
each other (for the case
\(\alpha\cdot \beta=0\)).
This is due to the fact that the weights of a minimal representation form a
single Weyl orbit. Thus we find that
\begin{equation}
    \dot{L}=[L,M]\Longleftrightarrow \dot{Q}=P,
   \quad \dot{P}=-m^2\left(\sum_{\alpha\in\Pi}\exp({\alpha\cdot Q})\alpha+
   \exp({\alpha_0\cdot Q})\alpha_0\right).
\end{equation}

\subsection{Root type Lax pair for simply laced root systems}

This type of Lax pairs applies universally
to all the Calogero-Moser models based on
simply laced root systems  \cite{bos, bcs2}. Its representation
space is the set of roots \(\Delta\) itself. Thus it is not related to any
representation of the associated algebra \({\mathfrak g}\) except for the
simplest case of \(A_1\).  This type of Lax pairs does {\em not} have  a
well-defined limit to the Lax pair of the corresponding Toda model.
Thus the Lax pair of a Toda model which is not  Lie algebra valued cannot be
constructed in this way.

The root type Lax pair for simply laced root systems reads:

\begin{eqnarray}
    L(q,p,\xi) & = & p\cdot H + X + X_{d},\quad \nonumber\\
    M(q,\xi) & = & D+Y+Y_{d}.
    \label{eq:genLaxform}
\end{eqnarray}
All of the matrices are labelled by the roots, \(\alpha, \beta, \gamma
\in\Delta\). The matrices \(H\), \(D\) and \(X\) have a similar structure to
those in the minimal type Lax pair, except that
\begin{displaymath}
E(\alpha)_{\beta\gamma}=\delta_{\beta-\gamma, \alpha}.
\end{displaymath}
The matrices $X_d$ and $Y_d$ are special for the root type Lax pair:
\begin{equation}
    X_d=2ig\sum_{\alpha\in\Delta}
    x_{d}(\alpha\cdot q, \xi)E_{d}(\alpha),\quad
    Y_d=ig\sum_{\alpha\in\Delta}
    y_{d}(\alpha\cdot q, \xi)E_{d}(\alpha),\quad
    E_{d}(\alpha)_{\beta \gamma}=\delta_{\beta-\gamma,2\alpha}.
    \label{eq:XYrdef}
\end{equation}
The functions \(x(u,\xi)\) and \(x_d(u,\xi)\) are solutions of  coupled
functional equations (II.2.24), (II.2.25), \cite{bos}.
They share similar properties. For example, similar to the factorisation of
 the potential in terms of \(x(u,\xi)\)  as in (\ref{fac}), we have
another factorisation
\begin{equation}
   x_d(u,\xi)x_d(-u,\xi)=-\wp(u)+\wp(2\xi).
   \label{fac2}
\end{equation}
Thus, for the consistent limit of the Lax pair, \(x_d(u,\xi)\) should have
the
same type of asymptotic behaviour as that of  \(x(u,\xi)\) (\ref{xcond}):
\begin{equation}
   gx_d(\alpha\cdot q,\xi)\to
   \left\{
   \begin{array}{cl}
      \mbox{finite},&
      \mbox{for}
      \quad
      \pm\alpha_i\in\Pi\quad  \mbox{and}\
      \pm\alpha_h,\\[6pt]
      0,&\mbox{otherwise.}
   \end{array}
   \right.
   \label{xdcond}
\end{equation}
However, this is not the case, since it is not compatible with the following
functional identity (III.3.21), \cite{bos}
\begin{equation}
   x(2u,\xi)x_d(-u,\xi)+x(-2u,\xi)x_d(u,\xi)=-\wp(u)+\wp(\xi),
   \label{secfac}
\end{equation}
which is a simple consequence of the general functional equation (II.2.25),
\cite{bos}.
We multiply \(g^2\) to (\ref{secfac}) and choose \(\alpha\) to be a simple
root
\[
   u=\alpha\cdot q=\alpha\cdot Q-2\omega_3\delta\rho\cdot\alpha,
   \quad \rho\cdot\alpha=1.
\]
Then the right hand side is finite as in (\ref{genlim2}). On the left hand
side
\[
   gx(\pm2\alpha\cdot q,\xi)\to0
\]
since both have twice the damping factor.
This means that either \(gx_d(\alpha\cdot q,\xi)\) or \(gx_d(-\alpha\cdot
q,\xi)\) or both must be divergent for simple roots \(\alpha\).
Thus the desired asymptotic behaviour  of  \(gx_d(u,\xi)\) (\ref{xdcond})
is not achieved and the Lax pair has no consistent limit.
In Appendix \ref{xlimforms} we list the asymptotic forms of \(x_d\) and
other
functions.

\subsection{Other root type Lax pairs without Toda limits}
\label{nonexist}
The above argument for the non-existence of  Toda limits for
Calogero-Moser Lax pairs, based as it is
on the non-existence of finite \(x\)
and
\(x_d\) functions for the simple roots, can be applied to other
models whose Lax
pairs contain the  root type Lax pair (\(x\) and \(x_d\) functions). Thus
the
following Lax pairs do not have a consistent Toda limit:
\begin{enumerate}
\item Root type Lax pair based on long roots for untwisted and twisted
\(B_r\)
model.
\item Root type Lax pair based on short roots for untwisted
and twisted \(C_r\).
model
\item Root type Lax pair based on long and short roots for untwisted and
twisted \(F_4\) model.
\item Root type Lax pair based on long and short roots for untwisted and
twisted \(G_2\) model.
\end{enumerate}

The set of the long roots of \(B_r\) is the same as the set of the roots of
\(D_r\). Thus the \(B_r\) Lax pairs based on the long roots
for the untwisted and
the twisted models (see section 4 of paper III, \cite{bos}) contain the
functions \(x\) and \(x_d\).

The set of  short roots of \(C_r\) is the same as the set of  roots of
\(D_r\). Thus the \(C_r\) Lax pairs based on the short roots for the
untwisted
model (see section 4 of paper III, \cite{bos}) contain the functions \(x\)
and \(x_d\).
The  \(C_r\) Lax pair based on the short roots for the twisted model
contains the twisted functions \(x^{(1/2)}\)
and \(x_d^{(1/2)}\). For these functions the twisted analogue of the
identity
(\ref{secfac}) reads
\begin{equation}
   x^{(1/2)}(2u,\xi)x^{(1/2)}_d(-u,\xi)
   +x^{(1/2)}(-2u,\xi)x^{(1/2)}_d(u,\xi)
   =-\wp(u|\{\omega_1,2\omega_3\})+f(\xi).
   \label{secfactw}
\end{equation}
This can be obtained from (III.5.10), \cite{bos} and \(f(\xi)\) is a \(\xi\)
dependent constant of integration. Thus we know that \(x^{(1/2)}(\alpha\cdot
q,\xi)\) and \(x_d^{(1/2)}(\alpha\cdot q,\xi)\) for $\pm$ simple
roots cannot be
finite at the same time by the  argument given in the previous section.
It should be stressed that this conclusion does not depend on a particular
choice of solutions  \(x^{(1/2)}\)
and \(x_d^{(1/2)}\), e.g. (III.5.15), \cite{bos},
since (\ref{secfactw}) is a consequence of the functional
equations. In fact, it also applies to a different set of solutions
\(x^{(1/2)}\)
and \(x_d^{(1/2)}\) given in Appendix \ref{othersol} which are equivalent to
those used in
\cite{DHPh2} for the Lax pair of the \(F_4\) model based on  short roots.

Since the algebra \(F_4\) is self-dual, \(F_4^\vee=F_4\), the sets of  long
roots and of short roots have the same structure as the set of roots for
\(D_4\). Thus the \(F_4\) Lax pair based on the long roots for the untwisted
and twisted models (see section 4 of paper III, \cite{bos}) contains the
functions
\(x\) and \(x_d\).
The \(F_4\) Lax pair based on the short roots for the untwisted model
contains
the functions \(x\) and \(x_d\) and the twisted models the twisted functions
\(x^{(1/2)}\) and \(x_d^{(1/2)}\).

The  algebra \(G_2\) is also self-dual, \(G_2^\vee=G_2\).
The sets of  long roots
and of short roots have the same structure as the set of  \(A_2\) roots.
The same argument as that for \(F_4\) applies to this case, although the
twisting is threefold.

\subsection{Minimal type Lax pair for non-simply laced root systems}
\label{minlaxnon}

This applies to the untwisted and  twisted \(B_r\) models based on the
spinor
representation and
 the untwisted and  twisted \(C_r\) models based on the vector
representation. These can be handled in a unified way.
The Toda limits of the Lax pairs exist only for the
untwisted models and for both choices \(v=\rho,\ \rho^\vee\).

First let us consider the untwisted potential case.
The Lax pair contains one function
\(x\) only. The choice \(v=\rho\) is suitable for the Toda
limit because in this case the minimum of \(\rho\cdot\alpha (={1/
2})\) is
achieved by the short simple  roots and the long simple roots have
\(\rho\cdot\alpha =1\).
Therefore the different scalings, as explained in (\ref{coupscal}), for
\(g_L\)  and \(g_S\) would produce a consistent Toda Lax pair. That is
\(g_L=m_Le^{\omega_3\delta}\) and
\(g_S=m_Se^{\omega_3\delta/2}/\sqrt{2}\) with \(\delta<{1/
h^{\vee}}\) \((={1/ h^{\vee}})\) for non-affine (affine)
theories.

For the choice \(v=\rho^\vee\), \(\delta<1/
h\) \( (=1/ h)\),
\( g_L=m_L\,e^{\omega_3\delta}\), \(
g_S=m_S\,e^{\omega_3\delta}/\sqrt2\),
the minimum of \(\rho^\vee\cdot\alpha\) is achieved by all the simple roots.
Thus the minimal type Lax pair has a finite limit and gives non-affine
\(B_r\)
(
\(C_r\)) or affine
\(B_r^{(1)}\)  (\(C_r^{(1)}\)) Toda models for the same parameter range as
in
the minimal type Lax pairs for the simply laced root systems.

The Lax pair for the twisted potentials contains two functions, \(x\) and
\(x^{(1/2)}\). These functions satisfy the relations (III.5.23), \cite{bos}
\begin{equation}
    x(u,\xi)x^{(1/2)}(-u,\xi)+x(-u,\xi)x^{(1/2)}(u,\xi)
    =-2\wp(u)+
   g(\xi),
   \label{xxhiden}
\end{equation}
in which \(g(\xi)\) is a \(\xi\)-dependent constant of integration.
Based on this formula one can show that the Toda model limit does not exist
for either choice of \(v=\rho\) or \(\rho^\vee\) as in the root type Lax
pair
cases.

\subsection{Root type Lax pair for non-simply laced root systems}
\label{rotlaxnon}

In this subsection the root type Lax pair for non-simply laced root systems
which are not treated in subsection \ref{nonexist} will be discussed.
These are the untwisted and twisted \(B_r\) models based on the short roots
and
the untwisted and twisted \(C_r\) models based on the long roots.
As has been pointed out in paper II, \cite{bos}, the root type Lax pair for
the \(C_r\) model based on the long roots is  equivalent to the minimal type
Lax pair.  We will therefore discuss  Toda model
limits of the \(B_r\) model Lax
pair based on the short roots.

First let us consider the untwisted \(B_r\) model. In this case the Lax pair
contains the function \(x\) for the long roots and \(x_d\) for the short
roots.
For the choice \(v=\rho^\vee\), the long roots and short roots are
treated equally. Thus the same argument based on (\ref{secfac}) applies and
the
Lax pair has no consistent Toda model limit.
For the choice \(v=\rho\), the simple short roots have  half  the
decreasing factor of the simple long roots. This requires
\( g_L=m_L\,e^{\omega_3\delta}\), \(
g_S=m_S\,e^{\omega_3\delta/2}/2\sqrt2\) for the finite potentials.
 By multiplying \(g_Lg_S\) to (\ref{secfac}), we obtain
\[
 g_Lx(2u,\xi)g_Sx_d(-u,\xi)+g_Lx(-2u,\xi)g_Sx_d(u,\xi)
 =-g_Lg_S(\wp(u)-\wp(\xi)).
\]
Suppose that \(g_Lx(\alpha\cdot q)\) and  \(g_Sx_d(\alpha\cdot q)\) are both
finite for the simple roots. Let us choose  \(u=\alpha\cdot q\) with
\(\alpha\) being a simple short root. The left hand side is then finite, but
the right hand side is divergent since \(g_S^2\wp(\alpha\cdot q)\) is finite
by assumption and the right hand side has an extra divergent factor
\(g_L/g_S\propto e^{\omega_3\delta/2}\). Thus  the Lax pair for the \(B_r\)
untwisted model based on the short roots has no consistent Toda model limit.

Next we consider the twisted model. In this case the Lax pair contains
the function \(x\) for the long roots and \(x_d^{(1/2)}\) for the short
roots.
For \(v=\rho\),  \(g_Lx(\alpha\cdot q,\xi)\) and
\(g_Sx_d^{(1/2)}(\alpha\cdot
q,\xi)\) are non-vanishing for the simple roots with the scalings
\(g_L=m_L\,e^{\omega_3\delta}\), \(
g_S=m_S\,e^{\omega_3\delta}/2\sqrt2\). We have a relation (III.5.22),
\cite{bos}
\begin{equation}
    x(2u,\xi)x_d^{(1/2)}(-u,\xi)+x(-2u,\xi)x_d^{(1/2)}(u,\xi)
    =-\wp(u|\{\omega_1,2\omega_3\})+\wp(\xi)-\wp(\omega_1),
   \label{xxdhf}
\end{equation}
which is consistent with the limit of the Hamiltonian.
The result is summarised as follows
( \(\alpha\) is a positive root):
\begin{equation}
   \begin{array}{llll}
      g_Lx(\alpha\cdot q,\xi)&\rightarrow&
      -m_L\exp({{\alpha\cdot Q}\over{2}})&
      \mbox{long simple roots}\\[6pt]
      &\rightarrow&{m_LZ}\exp(-{{\alpha\cdot
      Q}\over{2}})&\mbox{highest root}\\[6pt]
      &\rightarrow&0&\mbox{otherwise}\\[8pt]
      g_Sx_d^{(1/2)}(\alpha\cdot q,\xi)&
      \rightarrow&-m_S\exp({\alpha\cdot Q})&
      \mbox{short simple roots}\\[6pt]
      &\rightarrow&0&\mbox{otherwise}\\[8pt]
   \end{array}
\end{equation}
The resulting Lax pair is that of the \(C_r=(B_r)^\vee\) Toda model for
the parameter range
\[
   \delta(h-1)<\epsilon\leq1/2,\qquad \mbox{or}\quad \delta<1/h,\quad
   \epsilon=1/2,
\]
and the Lax pair of the Toda model based on the twisted affine
algebra \(A_{2r-1}^{(2)}=(B_r^{(1)})^\vee\)  for
\[
  \epsilon=1/2,\quad \delta=1/h.
\]

The finiteness of the potential for
 long and  short simple roots when \(v=\rho^\vee\)
requires different scalings of \(g_L\) and \(g_S\), namely
\(g_L=m_L\,e^{\omega_3\delta}\), \(
g_S=m_S\,e^{2\omega_3\delta}/2\sqrt2\), (\ref{coupscal}). For
\(\delta<{1/ h}\) \((={1/ h})\) we obtain
a finite non-affine (affine) Toda Lax  pair.

\subsection{Root type Lax pairs for the \(BC_r\)
model based on the long and/or
short roots}

The set of the middle roots of the  \(BC_r\)
root system is the same as the set of
 roots of
  \(D_r\). Consequently the root type Lax pair for  the \(BC_r\) model based
on
its middle roots does not have a consistent Toda model limit.
This leads us to consider the root type Lax pair based on the long or the
short
roots. Since the \(BC_r\) root system is self-dual, the Lax pair based on
the
long roots and that based on the short roots are related by the duality
transformation. Thus we consider only the root type Lax pair based on the
short roots. Our interest lies in the case in which the potentials for the
long, middle and short roots survive in the Toda limit. The other situations
are the same as in the \(B_r\) or \(C_r\) cases.

First let us consider the untwisted potentials. In this case the Lax pair
contains two functions, \(x\) for the long and the middle roots and
\(x_d\) for the short roots. For the choice of \(v=\rho^\vee\) given
in (\ref{bcrrho}), \(\rho^\vee\cdot\alpha=1\) for the simple middle and the
simple short roots. As shown in the root type Lax pair for the non-simply
laced case, \(g_Mx(\alpha\cdot q,\xi)\) and \(g_Sx_d(\alpha\cdot q,\xi)\)
cannot be simultaneously finite for the simple roots. Here the scalings of
the
couplings are determined by the Hamiltonian (\ref{gbcruntw}).
Thus the Lax pair does not have a Toda model limit for  untwisted
potentials.

Next we consider the most general twisted model, the extended \(BC_r\) model
with five independent coupling constants, (III.4.87), \cite{bos},
\cite{Ino2}.
We give the Lax pair for this model in Appendix \ref{extwbcr}.
It is easy to see that the extended function \(x_d^{(1/2)}\) for the
short roots cannot be finite at the same time as \(x^{(1/2)}\) for the
middle
roots.
Similarly, another  extended function \(x^{(1/2)}\) for the long roots
cannot be finite at the same time as \(x^{(1/2)}\) for the middle
roots. Thus for the consistent
Toda model limit we have to put \(g_{S_1}=g_{L_2}=0\)
and we are led to the ordinary twisted model with \(g_{L_1}\equiv g_L\),
\(g_M\) and
\(g_{S_2}\equiv g_S\) should be scaled as in (\ref{gbcrtw}), which is
determined by the limit of the Hamiltonian.
Now we have three  functions \(x\) for the long roots,
\(x^{(1/2)}\) for the middle roots and \(x^{(1/4)}_d\) for the short roots
in the Lax pair.
The coexistence of the function  \(x\) and
\(x^{(1/2)}\) cannot have a finite limit as discussed
in subsection \ref{minlaxnon}, see  (\ref{xxhiden}).
We thus  come to the conclusion that the \(A_{2r}^{(2)}\) Toda Lax pair
cannot be obtained as a limit from the Lax pair of the \(BC_r\)
Calogero-Moser Lax pair.

\bigskip
Let us close this section with a remark on the Toda model
limits of the Lax pairs of
Calogero-Moser model in the {\em generic} representations of the reflection
groups (see section five of \cite{bcs2}).
It is rather straightforward to see that the consistent
Toda model limit does not
exist in general.
The Lax pair contains the function
\(x(\alpha\cdot q,\alpha^\vee\!\cdot\mu\xi)\),
in which \(\mu\) is a generic basis vector of the representation.
This means that there are as many different functions of \(\alpha\cdot q\)
as
there are different values of \(\alpha^\vee\!\cdot\mu\) corresponding to one
potential function \(\wp(\alpha\cdot q)\). It is in general not possible
that
all of them have the same Toda model limit.

\section{Lax pairs of Toda models}
\setcounter{equation}{0}
In this section we give a general intuitive method of constructing
Lax pairs for Toda models. This method works for systems associated
with $A$, $B$, $C$ and $D$ series. The method works for those
root systems for which the simple roots and the affine root can be written
in the form $\pm 2e_i$,
$\pm e_i\pm e_j$ or $\pm e_i$, where $\{e_i\}$ forms an orthonormal basis.
A hint for this method actually comes  from the limits of the minimal
type Calogero-Moser Lax pairs described in the preceding section. We explain
the method with  simple examples.

\subsection{\(C_2^{(1)}\)}

First we discuss the case where the roots are
of the form $\pm 2e_i$ and $\pm e_i\pm e_j$  (the case of the form
$\pm e_i$ will be discussed later in the section). Let  us consider the
affine
$C_2^{(1)}$ Toda model. The Hamiltonian  is given by
\begin{equation}
   {\cal H}_{C_2}^{(1)}={1\over2}(p_1^2+p_2^2)+m^2e^{q_1-q_2}
   +{m^2\over2}\left(e^{2q_2}+e^{-2q_1}\right).
   \label{affc2ham}
\end{equation}
The canonical equations of motion are
\begin{eqnarray}
   \dot{p_1}&=&-{\partial{\cal H}\over{\partial q_1}}
   =-m^2e^{q_1-q_2}
   +m^2e^{-2q_1},
   \nonumber\\
   \dot{p_2}&=&-{\partial{\cal H}\over{\partial q_2}}
   =m^2e^{q_1-q_2}
   -m^2e^{2q_2},
   \label{affc2eq}
\end{eqnarray}
with $\dot q_1=p_1$ and $\dot q_2=p_2$.
In this case the two simple roots are $2e_2$ and $e_1-e_2$. The affine
root is given by $-2e_1$.
The $L$ matrix is constructed in the following way. The
dimension $dim$ of the Lax pair would be twice the rank of the
algebra
(this is not  always the case, as we remark afterwards).
The diagonal part of $L$ is given by,
\begin{equation}
   L_{i,i}=-L_{2r+1-i, 2r+1-i}=p_i,  \qquad i=1,2,..,r;~~r:\mbox{
   rank}.
\end{equation}
Now corresponding to every simple or affine root we have  off-diagonal
elements in $L$. Suppose we have a simple root of the form
$e_i-e_j$. Corresponding to
this we have ,
\begin{equation}
   L_{i,j}=-L_{j,i}=L_{2r+1-j, 2r+1-i}
   =-L_{2r+1-i,2r+1-j}=im\exp{1\over 2}(
   q_i-q_j),~~~i\neq j.
\end{equation}
For the present example $C_2$, $2r=4$ and there is a simple
root $e_1-e_2$ and correspondingly we have
\begin{equation}
   L_{1,2}=-L_{2,1}=L_{3,4}=-L_{4,3}=im\exp{1\over 2}(q_1-q_2).
\end{equation}
Another type of root is of the form $\pm2e_i$. Corresponding to these we
have counter diagonal elements in $L$ as follows:
\begin{equation}
   L_{i,2r+1-i}=-L_{2r+1-i,i}=\pm im\exp(\pm q_i).
\end{equation}
In the $C_2$ case corresponding to the affine root $-2e_1$ and a simple root
$2e_2$ then one has,
\begin{equation}
   L_{1,4}=-L_{4,1}=-im\exp(-q_1) ~~~{\rm and}~~~~
   L_{2,3}=-L_{3,2}=im\exp(q_2),
\end{equation}
respectively. The rest of the matrix elements of  $L$ vanishes. This
completes the construction of $L$ for affine
$C_2^{(1)}$ Toda model. Once $L$ is constructed $M$ can be written down very
easily in the following way. In $M$ the diagonal elements are vanishing,
\begin{equation}
   M_{i,i}=M_{2r+1-i,2r+1-i}=0,
\end{equation}
and the off-diagonal elements are determined by those of \(L\):
\begin{equation}
   M_{i,j}={i\over 2}|L_{i,j}|,~~~i\neq j.
\end{equation}
The matrices $L$ and $M$ for the affine $C_2^{(1)}$ model read,
\begin{equation}
L=\pmatrix{p_1 & ime^{(q_1-q_2)/2}& 0&-ime^{-q_1}\cr
           -i\,me^{(q_1-q_2)/2} & p_2 & i\,me^{q_2} &0\cr
            0 & -i\,me^{q_2} & -p_2 & i\,me^{(q_1-q_2)/2}\cr
            ime^{-q_1} & 0 &-i\,me^{(q_1-q_2)/2} &-p_1},
\end{equation}
\begin{equation}
M={1\over2}\pmatrix{0 & ime^{(q_1-q_2)/2}& 0&ime^{-q_1}\cr
           i\,me^{(q_1-q_2)/2} & 0 & i\,me^{q_2} &0\cr
            0 & i\,me^{q_2} & 0 & i\,me^{(q_1-q_2)/2}\cr
            ime^{-q_1} & 0 &i\,me^{(q_1-q_2)/2} &0}.
\end{equation}
It is easy to check
\begin{equation}
   \mbox{Tr}(L^2)=2(p_1^2+p_2^2)+4m^2e^{q_1-q_2}+2m^2\left(e^{2q_2}
   +e^{-2q_1}\right)=4{\cal
    H}_{C_2}^{(1)},
\end{equation}
and the Lax equation $\dot L=[L,M]$ is identical to the canonical equation
given in (\ref{affc2eq}).

\subsection{$A_3^{(2)}$ }
The second example is  the Toda model Lax pair
for the twisted affine algebra $A_3^{(2)}$. The simple roots of the affine
$A_3^{(2)}$ are identical to those of  $C_2$ already considered. The
affine root  is $-e_1-e_2$.
We adopt the following Hamiltonian:
\begin{equation}
   {\cal H}_{A_3}^{(2)}={1\over2}(p_1^2+p_2^2)+m^2\left(e^{q_1-q_2}
   +e^{-q_1-q_2}\right)+
   {m^2\over2}e^{2q_2}.
   \label{a32ham}
\end{equation}
The canonical equations of motion are
\begin{eqnarray}
   \dot{p_1}&=&-{\partial{\cal H}\over{\partial q_1}}=-m^2e^{q_1-q_2}
   +m^2e^{-q_1-q_2},
   \nonumber\\
   \dot{p_2}&=&-{\partial{\cal H}\over{\partial q_2}}=m^2e^{q_1-q_2}
   +m^2e^{-q_1-q_2}
   -m^2e^{2q_2}.
   \label{twia3eq}
\end{eqnarray}
For simple roots of the form $\pm(q_i+q_j)$ one writes elements in $L$ as
\begin{equation}
   L_{i,2r+1-j}=-L_{2r+1-j,i}=L_{j,2r+1-i}=-L_{2r+1-i,j}=\pm
   im\exp[\pm{1\over 2}(q_i+q_j)].
\end{equation}
So in this case corresponding to the affine root we have (the other terms
are the same as those in the $C_2$ case),
\begin{equation}
   L_{1,3}=-L_{3,1}=L_{2,4}=-L_{4,2}=-im\exp{1\over 2}(-q_1-q_2),
\end{equation}
and the Lax pair is a set of $4\times4$ matrices:
\begin{equation}
   L=\pmatrix{p_1 & ime^{(q_1-q_2)/2}& -ime^{(-q_1-q_2)/2}&0\cr
              -i\,me^{(q_1-q_2)/2} & p_2 & i\,me^{q_2}
   &-ime^{(-q_1-q_2)/2}\cr
               ime^{(-q_1-q_2)/2} & -i\,me^{q_2} & -p_2 &
   i\,me^{(q_1-q_2)/2}\cr
               0 &ime^{(-q_1-q_2)/2} &-i\,me^{(q_1-q_2)/2} &-p_1},
\end{equation}
\begin{equation}
   M={1\over2}\pmatrix{0 & ime^{(q_1-q_2)/2}& ime^{(-q_1-q_2)/2}&0\cr
              i\,me^{(q_1-q_2)/2} & 0 & i\,me^{q_2}
   &ime^{(-q_1-q_2)/2}\cr
               ime^{(-q_1-q_2)/2} & i\,me^{q_2} & 0 &
   i\,me^{(q_1-q_2)/2}\cr
               0 & ime^{(-q_1-q_2)/2} &i\,me^{(q_1-q_2)/2} &0}.
\end{equation}
It is easy to check
\begin{equation}
   \mbox{Tr}(L^2)=2(p_1^2+p_2^2)+4m^2e^{q_1-q_2}+
   4m^2e^{-q_1-q_2}+2m^2e^{2q_2}=4{\cal
    H}_{A_3}^{(2)}.
   \label{trlsqc2}
\end{equation}
In these two examples $L$ is hermitian and $M$  anti-hermitian.
Of course both of the Lax pairs constructed here are well-known.

\subsection{$D_3^{(2)}$ and $B_2^{(1)}$}
The third and fourth examples are the duals of those given in the previous
two
subsections. The dimensions of the Lax pairs for $B_2^{(1)}$ and $D_3^{(2)}$
are 5 and 6, respectively. These cases involve some of the simple roots of
the
form  $e_i$. First we construct the Lax pair for the $B_2^{(1)}$  Toda
model.
We could have written down the simple roots and the affine root of affine
$B_2^{(1)}$  as $e_1-e_2$, $e_2$ and
$-e_1-e_2$, respectively, by taking the co-roots of $D_3^{(2)}$,. In this
case
normalisation for the long roots is 2 whereas in the earlier parametrisation
it
is 4. For each simple root of the form $e_i$ one has to add an additional
row
and column to the Lax pair given above which is initially $2r$ dimensional
{\em i.e.} twice the rank. Since in
the $B_2$ case there is only one simple root
$e_2$ of this form the  dimension of the Lax pair is 5.
We mark the new row and column by
$0$, and we put the corresponding row and column in the middle of the Lax
pair matrices. The diagonal element $L_{0,0}=0$. Corresponding to $\pm e_i$
we
insert elements
\begin{equation}
   L_{i,0}=-L_{0,i}=L_{0,2r+1-i}=-L_{2r+1-i,0}=\pm im\exp(\pm{1\over 2}q_i).
\end{equation}
For the $B_2$ Toda model we then have
\begin{equation}
   L_{2,0}=-L_{0,2}=L_{0,3}=-L_{3,0}=im\exp({1\over 2}q_2).
\end{equation}
The rest of the elements of $L$ and $M$ are constructed in the usual manner
as
explained above and we have the following Lax pair:
\begin{equation}
   L=\pmatrix{p_1 & im_1e^{(q_1-q_2)/2}&0& -im_2e^{(-q_1-q_2)/2}&0\cr
              -i\,m_1e^{(q_1-q_2)/2} & p_2 & i\,me^{q_2/2}&0
   &-im_2e^{(-q_1-q_2)/2}\cr
      0& -i\,me^{q_2/2}&0&i\,me^{q_2/2}&0\cr
   im_2e^{(-q_1-q_2)/2} &0& -i\,me^{q_2/2} & -p_2 &
   i\,m_1e^{(q_1-q_2)/2}\cr
               0 &i\,m_2e^{(-q_1-q_2)/2}&0 &-i\,m_1e^{(q_1-q_2)/2} &-p_1},
\end{equation}
\begin{equation}
   M={1\over2}\pmatrix{0 & im_1e^{(q_1-q_2)/2}&0& im_2e^{(-q_1-q_2)/2}&0\cr
              i\,m_1e^{(q_1-q_2)/2} & 0 & i\,me^{q_2/2}&0
   &im_2e^{(-q_1-q_2)/2}\cr
      0& i\,me^{q_2/2}&0&i\,me^{q_2/2}&0\cr
   im_2e^{(-q_1-q_2)/2} &0& i\,me^{q_2/2} & 0 &
   i\,m_1e^{(q_1-q_2)/2}\cr
               0 &im_2e^{(-q_1-q_2)/2}&0 &i\,m_1e^{(q_1-q_2)/2} &0}.
\end{equation}
Once again  ${1\over 4}{Tr}(L^2)$  gives us the Hamiltonian for the
affine $B_2^{(1)}$ ($=C_2^{(1)}$) Toda model. Notice that
in the above we have introduced three coupling constants viz. $m_1$, $m_2$
and
$m$, but as shown in (\ref{coupredef})--(\ref{redefdham}) these are
irrelevant.
In a similar fashion we construct a 6 dimensional Lax pair for $D_3^{(2)}$
($=A_3^{(2)}$) model.
We write the simple and affine roots of $D_3^{(2)}$ as $e_2$, $e_1-e_2$ and
$-e_1$, respectively. The Lax pair reads
\begin{equation}
   L=\pmatrix{p_1 & i\,me^{(q_1-q_2)/2}&-i\,m_1e^{-q_1/2}&0& 0&0\cr
              -i\,me^{(q_1-q_2)/2} & p_2 &0& i\,m_2e^{q_2/2}&0 &0\cr
     i\,m_1e^{-q_1/2}&0&0&0&0&-i\,m_1e^{-q_1/2}\cr
    0&-i\,m_2e^{q_2/2}&0&0&i\,m_2e^{q_2/2}&0\cr
   0&0&0&-i\,m_2e^{q_2/2} & -p_2 & i\,me^{(q_1-q_2)/2}\cr
               0 &0 &i\,m_1e^{-q_1/2}&0&-i\,me^{(q_1-q_2)/2} &-p_1},
\end{equation}

\begin{equation}
   M={1\over2}\pmatrix{0 & i\,me^{(q_1-q_2)/2}&i\,m_1e^{-q_1/2}&0& 0&0\cr
              i\,me^{(q_1-q_2)/2} & 0 &0& i\,m_2e^{q_2/2}&0 &0\cr
     i\,m_1e^{-q_1/2}&0&0&0&0&i\,m_1e^{-q_1/2}\cr
    0&i\,m_2e^{q_2/2}&0&0&i\,m_2e^{q_2/2}&0\cr
   0&0&0&i\,m_2e^{q_2/2} & 0& i\,me^{(q_1-q_2)/2}\cr
               0 &0 &i\,m_1e^{-q_1/2}&0&i\,me^{(q_1-q_2)/2} &0}.
\end{equation}
Two of the coupling constants could be redefined and
the Hamiltonian could be written in terms of a single coupling constant
(mass
parameter),  for example $m$.

Before closing this section let us summarise that the intuitive method
gives  \(2r\) dimensional Lax pairs for \(C_r^{(1)}\) (vector
representation) and
\(A_{2r-1}^{(2)}\) and a \(2r+1\) dimensional (vector representation) Lax
pair
for
\(B_r^{(1)}\) and  a \(2r+2\) dimensional Lax pair for  \(D_{r+1}^{(2)}\)
Toda models.
Of course one can also construct the \(A_r^{(1)}\) and the \(D_r^{(1)}\)
Lax pairs in the vector representations in a similar fashion.

%%%%%%%%%%%%%%%%%%%%%%%%%%%%%%%%%
\section{Summary and discussions}

As shown in section \ref{potentials}, the Hamiltonians of the elliptic
Calogero-Moser models tend to those of Toda models as one of the periods
of the elliptic function goes to infinity, provided the dynamical
variables are properly shifted and the coupling constants are scaled.
Although both Calogero-Moser and Toda models are integrable, the
corresponding
limits of the Lax pairs are subtle.
This is partly because of the abundance of  Lax pairs,
{\em i.e.} there are many Lax pairs for a given Calogero-Moser model.
Here we list those Lax pairs of Calogero-Moser models which have Toda model
limits.

\begin{center}
\baselineskip=26pt
 \begin{tabular}{|c|c|c|c|} \hline \label{SubrootTbl}
   algebras & potential & Lax pair & Toda models\\ \hline
   $A_{r}$, $D_{r}$, $E_{6}$, $E_{7}$ & untwisted & minimal &${\mathfrak
g}$, ${\mathfrak g}^{(1)}$ \\
\hline
   $B_{r}$ & untwisted & minimal & $B_{r}$, $B_{r}^{(1)}$ \\ \hline
    $C_{r}$ & untwisted & minimal & $C_{r}$, $C_{r}^{(1)}$ \\ \hline
$B_{r}$ & twisted & short roots & $C_{r}$, $A_{2r-1}^{(2)}$ \\ \hline
$C_{r}$ & untwisted & long roots & $C_{r}$, $C_{r}^{(1)}$ \\ \hline
   \end{tabular}
\end{center}

\baselineskip=20pt
Here we give an intuitive explanation why some Calogero-Moser Lax pairs,
for example, the root type Lax pairs for simply laced root systems,
do not have Toda limits.
In some Lax pairs, there are two {\em different} functions, {\em e.g.}
\(x\) and \(x_d\), corresponding to the {\em same} potential in the
Hamiltonian.
These functions are related to the potentials by the factorisation
formulas, for example (\ref{fac}), (\ref{fac2}), and the limits of the
potentials are unique. That is, these two {\em different} functions must
have
the {\em same} limit for all the possible arguments \(\alpha\cdot q\)
belonging to the potential.
This  is not compatible with the functional equations they must satisfy,
thus the postulated limits to Toda models do not exist in such cases.

In \cite{DHPh2} it is claimed that the Lax pair for the twisted \(F_4\)
model based on the short roots has a Toda model limit.
In subsection \ref{nonexist} we have shown that such a limit does not
exist.
Let us follow their logic here.
They use three functions \(x\), \(x^{(1/2)}\) and \(x^{(1/2)}_d\)
in our notation, and their solution for \(x^{(1/2)}\) and \(x^{(1/2)}_d\)
corresponds to (\ref{althalf}) and (\ref{althalfd}) given in Appendix
\ref{othersol}.
They show that \(g_Sx_d^{(1/2)}(\alpha\cdot q,\xi)\) has a non-vanishing
finite limit for \(\pm\) simple short roots and at the same time
it is claimed that \(g_Sx^{(1/2)}(\alpha\cdot q,\xi)\) vanishes for
positive short roots.
This would clearly violate the factorisation relation (the half
period analogue of (\ref{fac})) unless
\(g_Sx^{(1/2)}(-\alpha\cdot q,\xi)\) diverge for positive short roots.
AS a result, the Lax pair does not have a Toda model limit.

Let us assume, according to their claim, that
\(g_Sx^{(1/2)}(\pm \alpha\cdot q,\xi)\) vanishes for short roots.
This would lead to another contradiction with the formula
Tr\((L^2)\propto {\cal H}\), without invoking the factorisation formulas.
On the right hand side the limit of the potential is well defined and
unique,
whereas on the left hand side the short root potential has two
different sources,
one from the function \(x_d^{(1/2)}\) and the other from \(x^{(1/2)}\).
The former has a finite limit while the latter contribution vanishes.
This means that the relation Tr\((L^2)\propto {\cal H}\) is broken
in the limiting process.

In \cite{DHPh2} a Toda model Lax pair for the twisted affine Lie algebra
\(D_{r+1}^{(2)}\) is obtained as a limit of a Lax pair for the twisted
\(C_r\) Calogero-Moser model in a \(2r+2\) dimension representation
\cite{DHPh1},
which does not belong to the minimal type or root type Lax pairs developed
in
our series of papers \cite{bos}.

\begin{center}
{\bf ACKNOWLEDGMENTS}
\end{center}
We thank A.\,J.\, Bordner for useful discussion and A.\,J.\,Pocklington
for reading and improving the text.
This work is partially supported  by a Grant-in-aid
from the Ministry of Education, Science and Culture,
Priority Area
``Supersymmetry and unified theory of elementary particles"
(\#707). S.\,P.\,K. is supported by the Japan Society
for the Promotion of Science.

\newpage
%%%%%%%%%%%%%%%%%%%%%%%
\begin{center}
{\LARGE\bf Appendix}
\end{center}
\appendix
\section{ Elliptic functions}
\label{ellipfuns}
\setcounter{equation}{0}

\renewcommand{\theequation}{A.\arabic{equation}}

Here we collect some  mathematical formulas which will be used in the
main text.
For the
elliptic functions we follow the notation and  conventions of
\cite{Erd} throughout this paper.
The Weierstrass
function
\(\wp\) is  a doubly periodic meromorphic function
with a pair of  primitive
periods
\(\{2\omega_1,2\omega_3\}\), \(\Im(\omega_3/\omega_1)>0\):
\begin{equation}
   \wp(u)\equiv\wp(u|\{
   2\omega_1,2\omega_3\})={1\over{u^2}}+
   \sum_{m,\,n}{}^\prime\left[{1\over{(u-\Omega_{m,\,n})^2}}-
   {1\over{\Omega_{m,\,n}^2}}\right],
   \label{periods}
\end{equation}
in which \(\Omega_{m,\,n}\) is a period
\[
   \Omega_{m,\,n}=2m\omega_1+2n\omega_3
\]
and \(\sum{}^\prime\) denotes the summation over all
integers, positive, negative and zero, excluding
\(m=n=0\). The Weierstrass sigma function \(\sigma(u)\) is
defined from
\(\wp(u)\) via the Weierstrass zeta function \(\zeta(u)\)
as:
\begin{eqnarray}
   \wp(u)&=&-\zeta^\prime(u),\qquad \quad
   \zeta(u)=d\log\,\sigma(u)/du=\sigma^\prime(u)/\sigma(u),
   \nonumber\\
   \zeta(u)&\equiv&\zeta(u|\{
   2\omega_1,2\omega_3\})={1\over{u}}+
   \sum_{m,\,n}{}^\prime\left[{1\over{u-\Omega_{m,\,n}}}
   +{1\over{\Omega_{m,\,n}}}+{u\over{\Omega_{m,\,n}^2}}\right],
   \nonumber\\
   \sigma(u)&\equiv&\sigma(u|\{
   2\omega_1,2\omega_3\})=u\,
   \prod_{m,\,n}{}^\prime\left(1-{u\over{\Omega_{m,\,n}}}\right)
   \exp\left[
   {u\over{\Omega_{m,\,n}}}+{u^2\over{2\Omega_{m,\,n}^2}}\right],
\end{eqnarray}
in which \(\prod{}^\prime\) denotes the product over all
integers, positive, negative and zero, excluding
\(m=n=0\).
For the parametrisation of the periods
\[
 \omega_1=-i\pi, \quad
   \omega_3 \in{\bf R}_{+},\quad
   \tau\equiv{\omega_{3}\over{\omega_{1}}}=i\omega_{3}/\pi,
\quad q=e^{i\tau\pi}=e^{-\omega_3}
\]
the following expansion formulas for the elliptic functions are
useful
\footnote{See, for example, \cite{Lawden} p221.}:
\begin{equation}
   \zeta(u)={i\eta_1u\over\pi}+{1\over2}\coth
   {u\over2}-2\sum_{n=1}^\infty{q^{2n}\over{1-q^{2n}}}\sinh nu,
   \label{zetaexp}
\end{equation}
\begin{equation}
   \wp(u)=({1\over{2\pi}})^2\left[-4i\eta_1\pi+{\pi^2\over{\sinh^2(u/2)}}
   +8\pi^2\sum_{n=1}^\infty{nq^{2n}\over{1-q^{2n}}}\cosh nu\right],
   \label{wpexp}
\end{equation}
in which
\begin{equation}
   \eta_1=\zeta(\omega_1)=i\pi\left({1\over{12}}-
   2\sum_{n=1}^\infty{nq^{2n}\over{1-q^{2n}}}\right)
   {{}\atop{{\longrightarrow}\atop{\omega_3\to+\infty}}}
   i{\pi\over{12}}.
\end{equation}
The exponential potentials can be obtained from the elliptic potentials
simply
by the following way. For a shifted argument
\begin{equation}
   u=U-\omega_3\ell,\quad |\ell|<1,
   \label{shiftu}
\end{equation}
the summations in (\ref{zetaexp}) and (\ref{wpexp}) can be neglected in the
limit
\(\omega_3\to+\infty\) to obtain
\begin{equation}
   \zeta(u)={i\eta_1u\over\pi}+{1\over2}\coth
   {u\over2},
   \label{zetaappr}
\end{equation}
\begin{equation}
   V_L(u)\approx {1\over{12}}+{1\over4}{1\over{\sinh^2u/2}},\quad
   V_S(u)\approx {1\over{3}}+{1\over{\sinh^2{u}}}.
   \label{wpappr}
\end{equation}
In other words we obtain the following asymptotic formulas:
\begin{equation}
   V_L(u)\approx e^{\pm U\mp \omega_3\ell}+const,\qquad
   V_S(u)\approx 4e^{\pm 2U\mp 2\omega_3\ell}+const,\quad
   \ell\grtrless0.
   \label{wpscal}
\end{equation}
Thus by appropriate scalings of the coupling constants
\begin{equation}
   g_L=m_L\,e^{\omega_3|\ell|/2},\quad
   g_S=m_S\,e^{\omega_3|\ell|}/2,\quad |\ell|\leq1,
\end{equation}
the exponential
potentials can be obtained:
\begin{equation}
   g^2_LV_L(u)\approx m^2_L\,e^{\pm U},\qquad
   g^2_SV_S(u)\approx m^2_S\,e^{\pm 2U},\quad
   \ell\grtrless0,\quad |\ell|<1,
   \label{gwpscal}
\end{equation}
in which we understand the constant parts are properly subtracted.
The extreme case \(|\ell|=1\) deserves special attention
\begin{equation}
   g^2_LV_L(U\pm\omega_3)\approx m^2_L\left(e^U+e^{-U}\right),\qquad
   g^2_SV_S(U\pm\omega_3)\approx 4m^2_S\left(e^{2U}+e^{-2U}\right).
   \label{g1wpscal}
\end{equation}
It should be remarked that any shift of the argument \(u\) of the potentials
which is proportional to \(\omega_3\) can always be written in the form
(\ref{shiftu}) with \(|\ell|\leq1\), due to the \(2\omega_3\) periodicity
of the \(\wp\) function.
Thus the shifted potential functions always decrease exponentially
\(e^{-\omega_3|\ell|}\), up to an additive constant term.
Only those potential terms which have the minimal decrease can be made
finite by appropriate scalings of the coupling constants.

The corresponding approximation formula for the \(\sigma\) function
reads
\begin{equation}
   \sigma(u)\approx 2\sinh{u\over2}\,\exp(-{u^2\over{24}}),
   \label{sigapp}
\end{equation}
in which \(u\) is shifted as in (\ref{shiftu}).
Since \(\sigma(u)\) is quasi-periodic in \(u\), we need two additional
asymptotic formulas corresponding to plus (minus) one period shift.
For
\begin{equation}
   u=u_0\pm2\omega_3,\quad u_0=U-\omega_3\ell,\quad |\ell|<1,
   \label{shiftu2}
\end{equation}
we have
\begin{equation}
   \sigma(u)\approx -2\sinh{u_0\over2}
   e^{\pm u_0+\omega_3}\,\exp(-{u^2\over{24}}).
   \label{sigapp2}
\end{equation}

\section{Other solutions of the functional equations}
\label{othersol}
\setcounter{equation}{0}

\renewcommand{\theequation}{B.\arabic{equation}}

Here we present without proof some new sets  of solutions
to the functional equations for the twisted
functions
 \(\{x^{(1/2)}, x_d^{(1/2)}\}\) and similar functions for the
\(G_2\) model,
\(\{x^{(1/3)}, x_d^{(1/3)}, x_t^{(1/3)}\}\).
These solutions are closely related to the ones given in the Appendix
of \cite{bcs2} and the derivation is similar.
For \(\{x^{(1/2)}, x_d^{(1/2)}\}\)
we have
\begin{equation}
   x^{(1/2)}(u,\xi)=\biggl[{\sigma(\xi-u)
   \over{\sigma(\xi)
   \sigma(u)}}
   -{\sigma(\xi-u-\omega_{1})
   \over{\sigma(\xi)
   \sigma(u+\omega_{1})}}
   \exp[\eta_{1}\,\xi]\biggr]
\label{althalf}
\end{equation}
and
\begin{equation}
   x_d^{(1/2)}(u,\xi)=\biggl[{\sigma(2\xi-u)
   \over{\sigma(2\xi)
   \sigma(u)}}
   +{\sigma(2\xi-u-\omega_{1})
   \over{\sigma(2\xi)
   \sigma(u+\omega_{1})}}
   \exp[2\eta_{1}\,\xi]\biggr].
\label{althalfd}
\end{equation}
Only the sign of the second term of (\ref{althalf})
is different from (A.27) of \cite{bcs2}.
For the \(G_2\) functions we have
\begin{eqnarray}
   x^{(1/3)}(u,\xi) &=& \biggl[{\sigma(\xi-u)
   \over{\sigma(\xi)
   \sigma(u)}}
   +\lambda{\sigma(\xi-u-{2\omega_{1}\over 3})
   \over{\sigma(\xi)
   \sigma(u+{2\omega_{1}\over
   3})}}\exp[(2/3)\eta_{1}\,\xi]  \nonumber\\
   &&\hspace{2cm} + \lambda^2{\sigma(\xi-u-{4\omega_{1}\over 3}
   )
   \over{\sigma(\xi)
   \sigma(u+{4\omega_{1}\over
   3})}}\exp[(4/3)\eta_{1}\,\xi]\biggr],
\label{altthr}
\end{eqnarray}
in which \(\lambda\) is a cubic root of unity,
\(\lambda=e^{\pm2i\pi/3}\) and
\begin{eqnarray}
   x_d^{(1/3)}(u,\xi) &=& \biggl[{\sigma(2\xi-u)
   \over{\sigma(2\xi)
   \sigma(u)}}
   +\lambda^2{\sigma(2\xi-u-{2\omega_{1}\over 3})
   \over{\sigma(2\xi)
   \sigma(u+{2\omega_{1}\over
   3})}}\exp[(4/3)\eta_{1}\,\xi]  \nonumber\\
   &&\hspace{2cm} + \lambda{\sigma(2\xi-u-{4\omega_{1}\over 3}
   )
   \over{\sigma(2\xi)
   \sigma(u+{4\omega_{1}\over
   3})}}\exp[(8/3)\eta_{1}\,\xi]\biggr],
\label{altthrd}
\end{eqnarray}
\begin{eqnarray}
   x_t^{(1/3)}(u,\xi) &=& \biggl[{\sigma(3\xi-u)
   \over{\sigma(3\xi)
   \sigma(u)}}
   +{\sigma(3\xi-u-{2\omega_{1}\over 3})
   \over{\sigma(3\xi)
   \sigma(u+{2\omega_{1}\over
   3})}}\exp[2\eta_{1}\,\xi]  \nonumber\\
   &&\hspace{2cm} + {\sigma(3\xi-u-{4\omega_{1}\over 3}
   )
   \over{\sigma(3\xi)
   \sigma(u+{4\omega_{1}\over
   3})}}\exp[4\eta_{1}\,\xi]\biggr].
\label{altthrt}
\end{eqnarray}
They satisfy (III.5.10) for \(n=3\) and (III.5.12), (III.5.13) \cite{bos}.

\section{Extended Twisted \(BC_r\) root system Lax
pair with five independent couplings based on short roots}
\label{extwbcr}
\setcounter{equation}{0}
\renewcommand{\theequation}{C.\arabic{equation}}
Here we write the root type Lax pair for the twisted $BC_r$ system
based on the short roots.

The pattern of short root-- short root  is:
\begin{equation}
   BC_r:\qquad \mbox{short root}
   - \mbox{short root}=\left\{
   \begin{array}{l}
      \mbox{long root}\\
      \mbox{middle root}\\
      2\times \mbox{short root}\\
      \mbox{non-root}
   \end{array}
   \right.
   \label{bcnshsh}
\end{equation}
Based on this one can construct the Lax pair as
\begin{eqnarray}
    L(q,p,\xi) & = & p\cdot H + X_m + X_{d}+X_l, \nonumber\\
    M(q,\xi) & = & D_m+D_L+Ds+Y_m+Y_{d}+Y_l,
    \label{eq:bcnshLaxform}
\end{eqnarray}
in which $X_m$ ($Y_m$) corresponds to short root $-$ short root=
middle root, $X_L$ ($Y_L$) corresponds to short root $-$ short root=
long root and $X_d$ ($Y_d$) corresponds to short root $-$ short root=
$2\times$ short root:
\begin{eqnarray}
    X_d&=&2i\sum_{\lambda\in\Delta_S}\left[
    g_{S_1}x_{d}^{(1/2)}(\lambda\cdot q,\xi)+
    g_{S_2}x_{d}^{(1/4)}(\lambda\cdot q,
    \xi)\right]E_{d}(\lambda),
    \nonumber\\
X_m&=&ig_{M}\sum_{\alpha\in\Delta_M}
    x^{(1/2)}(\alpha\cdot q,\xi)
    E(\alpha),
    \nonumber\\
 X_l&=&i\sum_{\Xi\in\Delta_L}\left[
    g_{L_1}x(\Xi\cdot q,\xi)+
    g_{L_2}x^{(1/2)}(\Xi\cdot q,
    \xi)\right]E(\Xi),
    \nonumber\\
    Y_d&=&i\sum_{\lambda\in\Delta_S}\left[
    g_{S_1}y_{d}^{(1/2)}(\lambda\cdot q,\xi)+
    g_{S_2}y_{d}^{(1/4)}(\lambda\cdot q,
\xi)\right]E_{d}(\lambda),
\nonumber\\
   Y_m&=&i g_{M}\sum_{\alpha\in\Delta_M}
   y^{(1/2)}(\alpha\cdot q,\xi)
   E(\alpha),
\nonumber\\
   Y_l&=&i\sum_{\Xi\in\Delta_L}\left[
    g_{L_1}y(\Xi\cdot q,\xi)+
    g_{L_2}y^{(1/2)}(\Xi\cdot q,
\xi)\right]E(\Xi),
\nonumber\\
    E_{d}(\lambda)_{\mu \nu}&=&\delta_{\mu-\nu,2\lambda},
\quad E(\alpha)_{\mu \nu}=\delta_{\mu-\nu,\alpha},
\quad E(\Xi)_{\mu \nu}=\delta_{\mu-\nu,\Xi}
    \label{eq:bncnshXYrdef}
\end{eqnarray}
\begin{equation}
    (Ds)_{\mu \nu}= \delta_{\mu,
\nu}(Ds)_{\mu},\quad
    (Ds)_{\mu}=i\left[
    g_{S_1}\wp(\mu\cdot q|\{\omega_1,2\omega_3\})+
    g_{S_2}\wp(\mu\cdot q|\{\omega_1/2,2\omega_3\})\right],
    \label{eq:bcnlndsHD1}
\end{equation}
\begin{equation}
    (D_m)_{\mu \nu}= \delta_{\mu,
\nu}(D_m)_{\mu},\quad
    (D_m)_{\mu}=ig_M\sum_{\alpha\in\Delta_M,\
    \alpha\cdot\mu=1}\wp(\alpha\cdot q|\{\omega_1,2\omega_3\}),
    \label{eq:bcnlndsHD2}
\end{equation}
\begin{equation}
    (D_l)_{\mu \nu}= \delta_{\mu,
\nu}(D_l)_{\mu},\quad
    (D_l)_{\mu}=i\left[
    g_{L_1}\wp(2\mu\cdot q)+
    g_{L_2}\wp(2\mu\cdot q|\{\omega_1,2\omega_3\})\right].
    \label{eq:bcnlndsHD3}
\end{equation}
%In these formulas \(\Delta_L\), \(\Delta_M\), \(\Delta_S\) are the sets
%of the long, middle and short roots, respectively.

%%%%%%%%%%%%%%%%%%

\section{Asymptotic forms of various functions appearing in the Lax pair}
\label{xlimforms}
\setcounter{equation}{0}
\renewcommand{\theequation}{D.\arabic{equation}}

In this section we give asymptotic forms of various functions, $x_d,~
x^{(1/2)},~x_d^{(1/2)},$ etc.
The dynamical variables and the spectral parameter are scaled as in
(\ref{qg-scale}) and (\ref{xiscale}) with
$\omega_1=-i\pi$ and
$\omega_3\rightarrow +\infty$.

\noindent
i) $x_d$ for positive roots \(\alpha\):
\begin{equation}
   \begin{array}{cccr}
      x_d(\alpha\cdot q,\xi)&\rightarrow&-1,&0<2\epsilon
      -\delta\rho\cdot\alpha<1,\\
      &\rightarrow&{Z^2}\exp(-{{\alpha\cdot
      Q}})\exp[\omega_3(2\delta \rho\cdot\alpha-4\epsilon)],&
      -1<2\epsilon -\delta\rho\cdot\alpha\leq0
   \end{array}
   \label{alimitgx}
\end{equation}
and for negative roots \(\alpha\):
\begin{equation}
   \begin{array}{cccr}
      x_d(\alpha\cdot q,\xi)&\rightarrow&\exp(-{{\alpha\cdot
      Q}})\exp[2\omega_3\delta
      \rho\cdot\alpha],\qquad& \qquad0<2\epsilon
      -\delta \rho\cdot\alpha<1,\\
      &\rightarrow&-{1\over{Z^2}}\exp[\omega_3(4\epsilon-2)],\qquad&
      \qquad 1\leq 2\epsilon -\delta \rho\cdot\alpha<2,
   \end{array}
   \label{alimitgx1}
\end{equation}
in which
\[
   x_d(u,\xi)={\sigma(2\xi-u)\over{\sigma(2\xi)\sigma(u)}}
   \exp({2\zeta(\xi) u}).
\]

\noindent
ii) $x^{(1/2)}$ for positive roots \(\alpha\):
\begin{equation}
   \begin{array}{cccr}
      x^{(1/2)}(\alpha\cdot q,\xi)&\rightarrow
      &-2\exp({3{\alpha\cdot Q}\over{2}})
      \exp[-3\omega_3\delta\rho\cdot\alpha],
      &0<\epsilon/2 -\delta\rho\cdot\alpha<1,\\
      &\rightarrow&{2Z}\exp(-{{\alpha\cdot
      Q}\over{2}})\exp[\omega_3(\delta \rho\cdot\alpha-2\epsilon)],
      &-1<\epsilon/2 -\delta\rho\cdot\alpha\leq0
   \end{array}
   \label{alimitgx2}
\end{equation}
and for negative roots \(\alpha\):
\begin{equation}
   \begin{array}{cccr}
      x^{(1/2)}(\alpha\cdot q,\xi)&\rightarrow&2\exp(-{{\alpha\cdot
      Q}\over{2}})\exp[\omega_3\delta
      \rho\cdot\alpha],&
      0<\epsilon/2 -\delta \rho\cdot\alpha<1,\\
      &\rightarrow &-{2\over{Z}}\exp({3{\alpha\cdot
      Q}\over{2}})\exp[\omega_3(2\epsilon-3\delta\rho\cdot\alpha-4)],&
      1\leq\epsilon/ 2 -\delta \rho\cdot\alpha<2,
   \end{array}
   \label{alimitgx3}
\end{equation}
in which
\[
 x^{(1/2)}(u,\xi) = {x(u,\xi/2)\,x(u+\omega_{1},\xi/2)\over
 x(\omega_{1},\xi/2)}\,\exp{[u(\zeta(\xi)-2\zeta(\xi/2))]},
\]
and \(x(u,\xi)\) is defined in (\ref{funx}).

\noindent
iii) $x_d^{(1/2)}$ for positive roots \(\alpha\):
\begin{equation}
   \begin{array}{cccr}
      x_d^{(1/2)}(\alpha\cdot q,\xi)&\rightarrow&-2\exp({{\alpha\cdot
      Q}})
      \exp[-2\omega_3\delta
      \rho\cdot\alpha], &0<\epsilon -\delta\rho\cdot\alpha<1,
      \\
      &\rightarrow&2{Z^2}\exp(-{{\alpha\cdot
      Q}})\exp[\omega_3(2\delta \rho\cdot\alpha-4\epsilon)],&
      -1<\epsilon -\delta\rho\cdot\alpha\leq0
   \end{array}
   \label{alimitgx4}
\end{equation}
and for negative roots \(\alpha\):
\begin{equation}
   \begin{array}{cccr}
      x_d^{(1/2)}(\alpha\cdot q,\xi)&\rightarrow&2\exp(-{{\alpha\cdot
      Q}})\exp[2\omega_3\delta
      \rho\cdot\alpha],&0<\epsilon -\delta \rho\cdot\alpha<1,\\
      &\rightarrow&-{2\over{Z^2}}\exp({{\alpha\cdot
      Q}})\exp[\omega_3(4\epsilon-2\delta\rho\cdot\alpha-4)],&
      1\leq\epsilon -\delta \rho\cdot\alpha<2,
   \end{array}
   \label{alimitgx5}
\end{equation}
in which
\[
x_d^{(1/2)}(u,\xi)={x(u,\xi)\,x(u+\omega_{1},\xi)\over
 x(\omega_{1},\xi)}.
\]
%%%%%%%%%%%%%%%%%%%%%%%%%%%%%%%%%%%%%%%%%%%%%%%%%%%%%

\end{document}